\documentclass{my_class}
\usepackage{graphics}
\usepackage{bm,amssymb,amsmath,graphicx}
\usepackage[dvipsnames]{color}
\usepackage[sort,authoryear]{natbib}
\bibpunct{(}{)}{,}{a}{}{,}
\newcount\ndots
\def\drawline#1#2{\raise 2.5pt\vbox{\hrule width #1pt height #2pt}}
\def\spacce#1{\hskip #1pt}
\def\solid{\drawline{24}{.5}\nobreak}
\def\bdash{\hbox{\drawline{4}{.5}\spacce{2}}}
\def\dashed{\bdash\bdash\bdash\bdash\nobreak}

\def\trian{\raise 1.25pt\hbox{$\scriptstyle\triangle$}\nobreak}
\def\squar{\raise 1.25pt\hbox{$\scriptstyle\Box$}\nobreak}
\def\dtrian{\raise 1.25pt\hbox
   {$\scriptscriptstyle\bigtriangledown$}\nobreak}
\def\circle{$\circ$\nobreak}

\def\solidcircle{$\bullet$\nobreak}

\newcommand{\myfig}[3]{\resizebox{#1}{#2}{\includegraphics*{#3}}}
\newcommand{\mylab}[3]{\raisebox{#2}[0mm][0mm]{%
\makebox[0mm][l]{\hspace*{#1}#3}}}

\begin{document}

\thispagestyle{empty}

\input psfig.sty

\jname{Annu. Rev. Fluid Mech.}
\jyear{2012}
\jvol{\bf 44}
\ARinfo{1056-8700/97/0610-00}

\title{\vspace{-45pt}The Significance of Simple Invariant Solutions in Turbulent Flows}

\markboth{Kawahara$\cdot$Uhlmann$\cdot$van Veen}{The Significance of Simple Invariant Solutions in Turbulent Flows}

\author{
Genta Kawahara
\affiliation{Graduate School of Engineering Science, Osaka University, Toyonaka, Osaka 560-8531, Japan;
email: kawahara@me.es.osaka-u.ac.jp}
Markus Uhlmann
\affiliation{
    Institute for Hydromechanics, %
    Karlsruhe Institute of Technology, %
    76128 Karlsruhe, %
    Germany;
email: uhlmann@kit.edu
}
Lennaert van Veen
\affiliation{Faculty of Science, University of Ontario Institute of Technology, Oshawa ON L1H 7K4, Canada;
email: Lennaert.vanVeen@uoit.ca
}
}


\begin{abstract}
Recent remarkable progress in computing power and numerical analysis
is enabling us to fill a gap in the dynamical systems approach to turbulence.
One of the significant advances in this respect has been the numerical discovery of simple
invariant sets, such as nonlinear equilibria and periodic solutions, in
well-resolved 
Navier--Stokes flows. 
This review describes some fundamental and practical aspects of dynamical systems theory for the
investigation of turbulence, focusing on recently found invariant solutions and their significance for the dynamical
and statistical characterization of low-Reynolds-number turbulent flows. It is shown that the
near-wall regeneration cycle of 
coherent structures
can be reproduced
by such solutions. The typical similarity laws of turbulence, i.e. the Prandtl wall law and the
Kolmogorov law for the viscous range, as well as the pattern
and intensity of turbulence-driven secondary
flow in a square duct can also be represented by these simple invariant solutions.
\end{abstract}

\maketitle

\newpage
\section{INTRODUCTION}

In 1983
the IUTAM (International Union of Theoretical and Applied Mechanics) Symposium on
Turbulence and Chaotic Phenomena in Fluids was held in Kyoto, Japan.
At that symposium a French participant showed a transparency on which
a cartoon was drawn of
two corps ready to do battle against each other
\citep[see][]{yamada02}.
One was the 
``traditional statistical theory'' army which claimed
that turbulence should be considered statistically
while the other was 
``chaos'' army which insisted that
turbulence could be described as deterministic chaos in terms
of dynamical systems theory.
In that period the new concept of chaos was starting to have a strong impact
on a variety of disciplines in science and engineering, including
turbulence research.
The above anecdote is an indication of
the expectation
that dynamical systems theory
would be a counterpart of the statistical theory
of turbulence, although there was a wide gap between
the low-dimensional chaotic systems discussed at that time and turbulent flow.
Recently, however, nonlinear equilibrium and periodic solutions
to the full Navier--Stokes equation have been found numerically in several flow
systems, and it has turned out that they are significant for
a theoretical description of not only transition to turbulence but
also fully turbulent flow.
This paper is devoted to a review of recently reported
dynamical systems approaches to the problem of low-Reynolds-number turbulence
based on such equilibrium and periodic solutions.
Several review articles have recently appeared concerning the dynamical
systems approach to the transition to turbulence \citep[see][]{kerswell05, eckhardt07, eckhardt08}.
Here we focus our attention on fully turbulent flow.

Although apparently
infinite-dimensional, the turbulent motion of a viscous fluid governed by
the Navier--Stokes equations, i.e., by a set of partial differential
evolution equations in a finite domain, is a finite-dimensional dynamical
system. Physically, the reason is that small-scale motions are smoothed
by viscosity.
There is a mathematical proof of the existence of approximate
inertial manifolds for the Navier--Stokes equations that provides a 
relation between the dominant modes and the higher-order
ones that can be truncated
\citep[see, e.g.][]{foias01}.
Within such a manifold,
the Navier--Stokes equations can be approximated by a finite-dimensional
system of ODEs (Ordinary Differential Equations).
For example,
\cite{keefe92} demonstrated numerically that a spatially periodic turbulent
channel flow
simulated with a spatial resolution of
$16\times 33\times 16$ at the friction Reynolds number $Re_\tau=80$
was really confined to a finite-dimensional strange
attractor whose Kaplan--Yorke dimension was estimated to be
$780$.
The Kaplan--Yorke dimension specifies dimension of a volume element
in phase space which neither expands nor shrinks on average over time.

Dynamical systems theory tells us that coherent structures in turbulent
flows may be thought of as low-dimensional invariant sets in phase space,
in the neighbourhood of which the system spends a substantial fraction of
time, as suggested 
by \cite{Jimenez87}. 
Spatio-temporally organized
structures appear when a turbulent state approaches such an invariant set.
Possible invariant solutions are simple saddles in phase space, such as
steady travelling waves, or time-periodic solutions. 
It is known that the statistical properties of chaotic
low-dimensional dynamical systems can be estimated from the coherent states
represented by unstable periodic orbits
\citep{cvitanovic88,artuso90a,artuso90b,christiansen97}.
In this review,
we will restrict ourselves to such relatively simple invariant solutions,
even though
coherent structures can also correspond to more complex sets, such as the
strange attractors of some highly reduced dynamical systems obtained from
the Navier--Stokes equations by proper orthogonal decomposition
\citep[see, e.g.][]{holmes96}.

The fundamentals of dynamical systems theory
are described in Section~\ref{dynsys}, including recent work on
computational algorithms.
In Section~\ref{review}
we briefly review recently found simple invariant solutions
for wall-bounded shear flows, such as
plane Couette, plane Poiseuille,
Hagen--Poiseuille and square-duct systems,
and for triply periodic flow, namely
Kida--Pelz flow
\citep{kida85}.
We discuss the significance of the invariant solutions
in dynamical and statistical descriptions of low-Reynolds-number turbulent flows
in Section~\ref{significance}.
The paper is concluded with an outlook for the
dynamical system approach to turbulence and its open 
issues
in Section~\ref{future}.

\section{DYNAMICAL SYSTEMS\label{dynsys}}

As argued above, viscous flow can be approximated by a system of ODEs.
The number of degrees of freedom this system comprises
determines the spatial resolution of the 
flow simulation. 
In order to simulate the flow accurately,
we must take into account all spatial scales that carry significant energy
and those which are responsible for its dissipation. 
It is known phenomenologically that
the number of resolution elements grows with the Reynolds number $Re$
as $Re^{9/4}$ \citep{landau}.
For three-dimensional simulations
of even weakly turbulent flows at $Re\gtrsim 10^2$ 
we should work on systems of ODEs with upward from $10^5$ degrees of freedom.
Here, we will discuss some of the theory and
practice of analysing such systems.

In order to apply dynamical systems theory to fluid
flow simulations, we reformulate the governing
equations as
\begin{equation}
\frac{\mbox{d}\bm{x}}{\mbox{d}t}=\bm{f}(\bm{x},\bm{p}),
\label{dynsys_1}
\end{equation}
where $\bm{x}$
is the vector of dependent variables, for instance Fourier or Fourier-Chebyshev
coefficients or components of the velocity at grid points. The vector field
$\bm{f}$ stems from
the discretized Navier--Stokes equation and depends on parameters $\bm{p}$. 
In the context of the current review, these parameters determine 
the kinematic fluid viscosity $\nu$,
the strength of forcing and the geometry of the flow domain. The vector $\bm{x}$ lives in a 
``phase space''
of dimension $N$, which usually coincides with $\mathbb{R}^N$.
We denote the flow of this system of ODEs as $\bm{\phi}(\bm{x},t;\bm{p})$, such that
for any fixed $\bm{x}$, $\bm{\phi}(\bm{x},t;\bm{p})$ is the solution which passes through
point $\bm{x}$ at $t=0$. 
The Jacobian
$\bm{J}=\{\partial f_i/\partial x_j(\bm{x};\bm{p})\}$ ($1\leq i,\,j\leq N$)
plays a central role in 
the analysis of stability of solutions through the linearised equations
\begin{equation}
\frac{\mbox{d}\bm{A}}{\mbox{d}t}=\bm{J}\bm{A} \ \text{(matrix form), or}\ \frac{\mbox{d}\bm{v}}{\mbox{d}t}=\bm{J}\bm{v}
\ \text{(vector form)},
\label{dynsys_2}
\end{equation}
where $\bm{v}\in\mathbb{R}^N$ is a perturbation vector, i.e. a column vector of the fundamental matrix $\bm{A}$.
For discretized Navier--Stokes
equations, we typically have access only to a single perturbation vector in a simulation, either through 
a modification of the DNS (Direct Numerical Simulation)
code to integrate the vector form of linearized equations (\ref{dynsys_2}) or 
through finite differences. 

A practical remark is in place here. It can be quite a challenge to recast legacy software for the
DNS of turbulence such that the vectors $\bm{x}$, $\bm{f}$ and $\bm{p}$ can
easily be manipulated. 
The truly independent variables $\bm{x}$ may have to be extracted from
larger data structures taking into account boundary conditions and
possible redundancy.
Recently, DNS software has been written with an 
intuitive interface between the 
DNS and the dynamical systems formulation \citep[see, for instance, the channelflow project by][]{gibson}.

\subsection{Invariant Solutions\label{invsets}}

An important question that dynamical systems theory addresses concerns the long term
behaviour of solutions to ODEs (\ref{dynsys_1}).
Chaotic dynamics of a system can be decomposed into building blocks,
namely simple invariant solutions and
their stable and unstable manifolds.

The simplest invariant solution is an equilibrium, at which  
$\bm{f}(\bm{x},\bm{p})=0$. Equilibria come in families, denoted by $\bar{\bm{x}}(\bm{p})$, which
persist under changes of the parameters as long as the Jacobian, $\bm{J}(\bar{\bm{x}},\bm{p})$, is not singular.
Along a family, an eigenvalues of $\bm{J}(\bar{\bm{x}},\bm{p})$ may cross
zero. In this case a saddle-node bifurcation occurs, at which 
two equilibria collide and disappear. If $\bm{J}(\bar{\bm{x}},\bm{p})$ has a pair of complex eigenvalues with zero real part,
a Hopf bifurcation occurs, at which a periodic solution is created.
Examples of both types of bifurcation were found in plane Couette flow by \citet{clever97}.
If, and only if, the real part of all eigenvalues is negative, the equilibrium is stable. Nearby initial conditions
will converge to it. A famous example is the laminar equilibrium in plane Couette flow.
If the real part of some eigenvalues is positive, the equilibrium is a saddle. 
Examples of saddle-type equilibria are three-dimensional equilibrium states in
plane Couette flow \citep{Nagata90,clever97}.

The simplest time-dependent invariant solution is periodic.
Any point $\bm{x}'$ on a periodic solution satisfies
$\bm{\phi}(\bm{x}',T;\bm{p})=\bm{x}'$
and the smallest positive $T$ for which this holds is called the period. 
Just like equilibria, periodic solutions come in families.
Bifurcations and stability of such a family are determined by the Floquet multipliers,
which are the eigenvalues of the monodromy matrix
$\bm{M}=\{\partial\bm{\phi}_i/\partial x_j(\bm{x}',T;\bm{p})\}$ ($1\leq i,\,j \leq N$),
i.e. the time-$T$ solution of the linearized equations (\ref{dynsys_2}).
There is always one multiplier equal to unity, corresponding
to perturbations along the periodic solution. 
A saddle-node bifurcation occurs if there is a 
second multiplier equal to unity. This happens, for instance, to the periodic solutions
in plane Couette flow, identified by \citet{KawKida01}, when continued
in the Reynolds number \citep{kawa3}. A periodic solution is called a saddle
if some of the multipliers are inside the unit circle in the complex plane and some are outside. 
The dynamical systems picture of turbulence suggests that infinitely
many saddle-type periodic solutions exist and the state of the system wanders between them irregularly.
This picture can be traced back all the way to Hopf \citep[][Appendix A]{cvita}. 

\subsection{Connecting Orbits\label{connect}}

Saddle-type invariant solutions regulate the global dynamics of a system through 
their stable and unstable manifolds. These manifolds
are defined as the sets of initial conditions which converge to 
the saddles in positive
and negative time, respectively \citep{wigg}. 
At the equilibria and periodic solutions, the (un)stable 
manifold coincides with the subspace spanned by the (un)stable eigenvectors of the Jacobian
and the monodromy matrix, respectively.

When a stable manifold intersects with
an unstable manifold, the intersection contains at least one solution curve,
called ``homoclinic'' if both manifolds are attached to the same saddle and 
``heteroclinic'' if they are attached to different saddles. Such connecting
orbits provide the pathways for the state of the system to wander from one saddle to another. 
Thus, with the dynamical systems picture of turbulence in mind,
it is important to study connecting orbits.
Computationally, this is a hard task. What we can decide purely on the basis of the stability
properties of the saddles is the codimension of a connecting
orbit, i.e. the number of parameters we need to tune in order for a connecting orbit to exist.
The codimension $c$ of an orbit connecting 
a saddle with an unstable manifold of dimension $d_{\rm u}$ to 
one with a stable manifold of dimension $d_{\rm s}$ is 
$c=N-d_{\rm u}-d_{\rm s}+1$.
For example, for a homoclinic orbit to an equilibrium we have 
$d_{\rm u}+d_{\rm s}=N$
so $c=1$, while for a homoclinic orbit to a periodic solution we have
$d_{\rm u}+d_{\rm s}=N+1$ so $c=0$. This means that we can only see orbits homoclinic
to equilibria if we vary one system parameter, but we can expect to find
orbits homoclinic to periodic solutions in a system with all parameters fixed.

\subsection{Computation and Continuation of Invariant Solutions\label{numcomp}}

The best available method for computing saddle-type invariant solutions is Newton iteration
because of its quadratic convergence. Given an approximate equilibrium $\bar{\bm{x}}_i$ we find
a better approximation after the Newton step
\begin{equation}
\bar{\bm{x}}_{i+1}=\bar{\bm{x}}_i+\delta \bm{x},\ \text{where}\ \bm{J}(\bar{\bm{x}}_i,\bm{p})\,\delta \bm{x}=-\bm{f}(\bar{\bm{x}}_i,\bm{p})
\label{dynsys_8}
\end{equation}
The obvious difficulty is that we need to construct, store and factorise the $N\times N$ Jacobian
to solve this linear system.
For periodic solution computations,
the situation is even worse, since the fundamental matrix $\bm{A}$ in the Newton step
\begin{equation}
\bm{x}'_{i+1}=\bm{x}'_{i}+\delta \bm{x},\ \text{where}\ [\bm{A}(\bm{x}'_i,T_i)-\mathbb{I}]\,\delta \bm{x}=-\bm{\phi}(\bm{x}'_i,T_i)+\bm{x}'_i
\label{dynsys_9}
\end{equation}
takes $N$ integrations over time interval $T_i$ to compute.
Although some of the early computations
were done using this approach \citep{kawa3,veen06}, 
in order to attain high spatial resolution 
it is necessary 
to switch to inexact linear solvers. Such methods construct an
approximate solution to Newton step (\ref{dynsys_8}) or (\ref{dynsys_9}) on the basis of of matrix-vector products,
for the computation of which it suffices to
integrate linearized equations (\ref{dynsys_2}) in the vector form.
Most commonly used are Krylov subspace methods, in particular GMRES \citep{saad}. 
The efficiency of this methods depends on the eigenvalue structure of the
matrix. If the eigenvalues are unfavourable, we need to precondition the linear
problem \citep[see, e.g.][for an example in Navier--Stokes flow]{tuck}. For periodic solutions,
the linear problem is generally
well conditioned. \citet{sanch2} give an estimate of the number of necessary
GMRES iterations in terms of bounds on the spectrum of $\bm{M}$.
In their example computations,
and many subsequent applications \citep[e.g.,][]{viswanath07}, the number of GMRES iterations is less than
100 for $N=\mbox{O}(10^5)-\mbox{O}(10^6)$.

Continuation of invariant solutions in parameters takes only a small 
extension of the Newton--Krylov algorithm \citep[see, e.g.][]{sanch2}. The same Newton--Krylov
method and preconditioner (if any) can be used \citep{dickson}.

In the discussion above we have tacitly assumed that the dynamical system (\ref{dynsys_1}) has no symmetries.
A discussion of symmetry, often important in turbulence,
is outside the scope of the current review. One important notion is that of relative equilibria (or travelling waves) and periodic 
solutions. These are observed as regular equilibria and periodic solutions in the appropriate moving
coordinate frame, and their bifurcation theory and numerical treatment are analogous to those described
above. A good starting point for the interested reader is \citet{simi}.

\subsection{SIDEBAR: Detection of Unstable Invariant Sets\label{detect}}

Newton-Krylov iteration for the computation
of invariant solutions
requires an approximate solution to start with. Three methods for obtaining
initial data have been used successfully for turbulent flows.

The first is one of bisection and generated most of the streak-dominated solutions discussed
in Section \ref{comparison}. In flows with a stable laminar equilibrium, we consider two initial
conditions, one leading to long-lived turbulence and one to quick laminarization.
Bisection of initial conditions between these points yields orbits which
closely approach near-laminar invariant solutions.

The second method is that of homotopy, and generated many of the solutions 
described in Section \ref{equilibria}.
An extra parameter is introduced in
the system, for instance the amplitude of a body force.
Invariant solutions may be found when changing this parameter and
traced back to the original system by continuation.

The third method is one of filtering. Approximate recurrences or equilibria are filtered
from a turbulent time series. Initial data thus obtained are usually coarse
and necessitate globally convergent Newton methods.
\citet{dennis} describes several variants which can be integrated with Krylov subspace methods. 
\citet{lan} formulated a global Newton
step for periodic solutions. Using their idea,
\citet{faze} computed periodic solutions in 
triply-periodic weak
turbulence.

\section{SIMPLE INVARIANT SOLUTIONS\label{review}}

Even using Krylov subspace algorithms, the numerical computation of simple invariant solutions
in a large domain with inflow and outflow
boundary conditions comparable to experiments
is beyond our computational
capabilities at this moment.
However, 
DNS of turbulent states in
periodic domains 
is
known to reproduce the statistics of
experiments well
\citep[see, e.g.][]{jimenez11}.
It is also known that
at least
part of the buffer layer
is well represented
even by
the minimal flow units
which \cite{jmoin}
have 
determined
by
minimizing the streamwise and spanwise dimensions
of a computational periodic
box for plane Poiseuille turbulence.
The same approach was later applied
to plane Couette flow by \cite{Hamilton95}.
Such reduced systems are small enough for their
coherent structures to be described in terms of 
simple invariant
solutions.
In this section we see that several 
of these
have been found for doubly-
or singly-periodic minimal domains in plane Couette, Poiseuille,
Hagen--Poiseuille and square-duct flows,
and for triply-periodic Kida--Pelz flow.

\subsection{Steady and Travelling-Wave Solutions\label{equilibria}}

\begin{figure}[t]
\vspace*{3mm}
\centerline{%
\myfig{!}{0.36\textwidth}{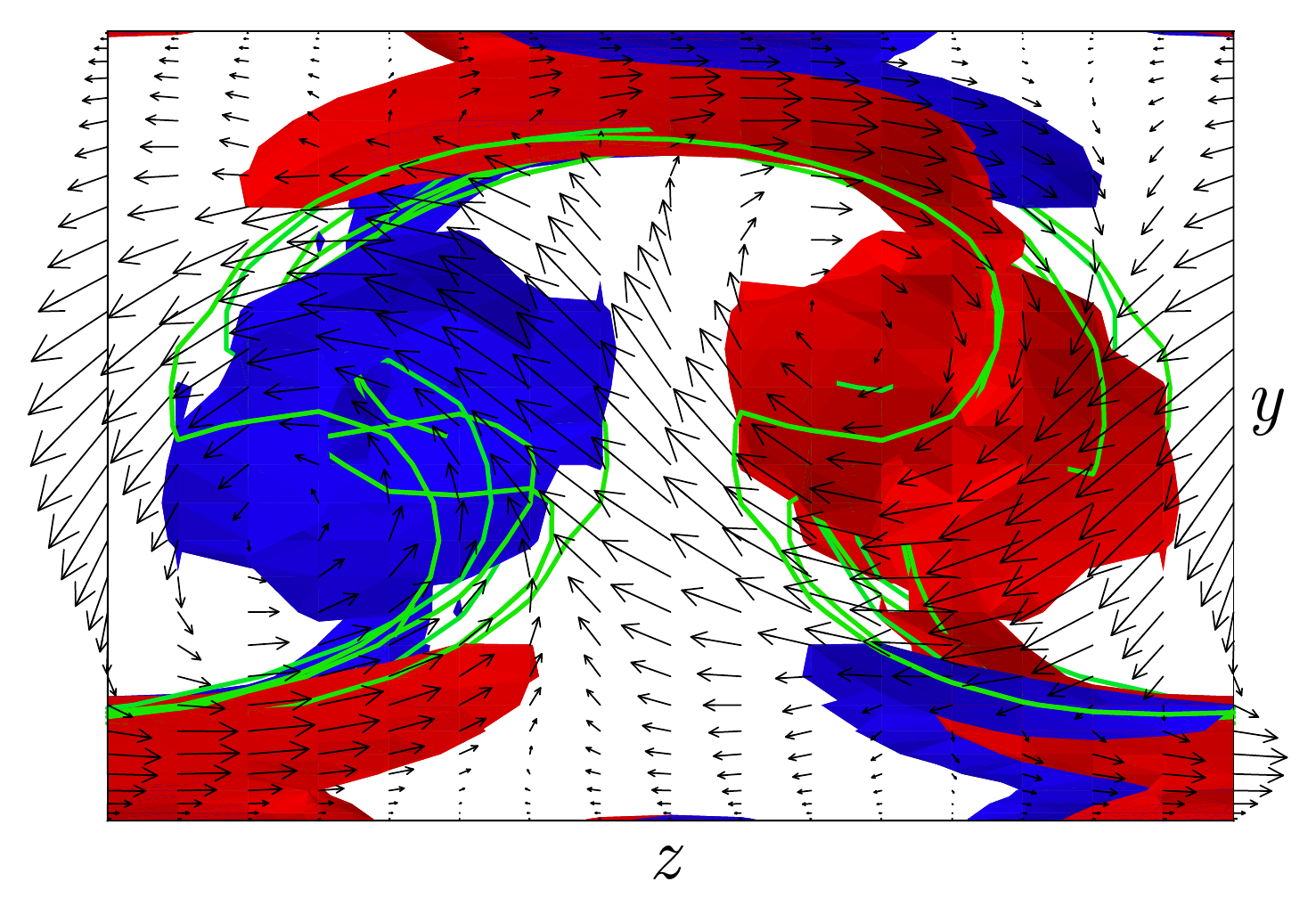}%
\mylab{-0.28\textwidth}{0.37\textwidth}{{\bf a}}
\myfig{!}{0.36\textwidth}{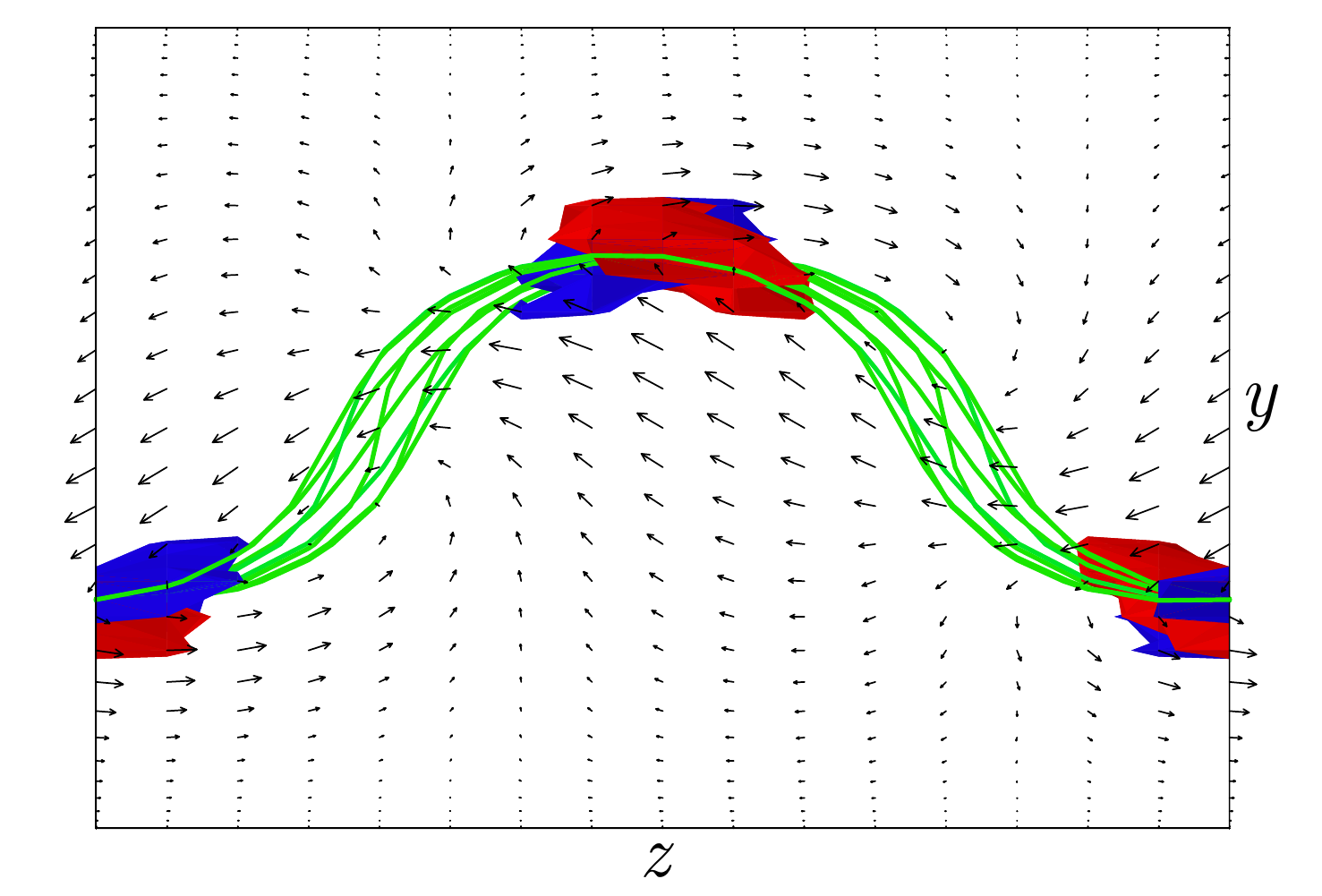}%
\mylab{-0.28\textwidth}{0.37\textwidth}{{\bf b}}
}
\caption[]{%
Projection of the streamwise ($x$) vorticity $\omega_x$ and velocity fields on the
cross-stream plane $(z,y)$ of the upper- and lower-branch \cite{Nagata90}
solutions for plane Couette flow 
recomputed by \cite{jim05}
at Reynolds number
$Re\equiv U h/\nu=400$,
$U$ and $h$ being
half the difference of the two wall velocities and
half the wall separation, respectively.
The streamwise and spanwise period for the solutions are
$L_x/h\times L_z/h =2\pi\times 0.9\pi$.
The green lines are
different sections of the surface of null streamwise velocity $u=0$, i.e, the critical layer,
and their corrugation represents a low-velocity streak.
Arrows are the cross-plane
velocities at $x=0$.
The red and blue objects are isosurfaces
$\omega_x^+\equiv \omega_x\nu/u_\tau^2=\pm 0.155$
representing streamwise vortices
($u_\tau$ being the wall friction velocity).
({\it a}) upper branch, $L_z^+\equiv L_z u_\tau/\nu=99$;
({\it b}) lower branch, $L_z^+=67$.
}
\label{fig:nagata}
\end{figure}

The first solutions of this kind were obtained by \cite{Nagata90} in plane
Couette flow. That flow is stable to infinitesimal disturbances at all
Reynolds numbers, so that nonlinear solutions cannot simply be found by
continuation from the laminar state. \cite{Nagata90} found his solution by
the homotopy method,
imposing a spanwise rotation on the system, which led to a sequence of
bifurcations of two- and three-dimensional steady solutions
from the laminar state. He then extended one of those three-dimensional nonlinear
solutions to the non-rotating case. \cite{clever92} obtained the same
three-dimensional solution by imposing a temperature difference
between the two horizontal walls of a Couette flow, and the same solution
was again found by \cite{Wally98,Wally03} who
imposed an artificial body force (see Section \ref{cycle}).
The three-dimensional equilibrium
solution found by those three groups arises from a saddle-node bifurcation
at a finite value of the Reynolds number, above which it splits into two
solution branches.
As shown in Figure~\ref{fig:nagata},
the solutions of the upper branch
contain
a wavy low-velocity streak
flanked with
a pair of staggered counter-rotating
streamwise
($x$)
vortices.
The solutions in the lower branch are
roughly streamwise independent and
closer to the laminar state. Other equilibrium solutions have been found more
recently for Couette flow
\citep{nagata97,schmiegel99,gibson08,gibson09,itano09}. They are not
necessarily related to the one originally identified by \cite{Nagata90},
but they have been used to discuss coherent structures 
or
the subcritical transition to turbulence.

\begin{figure}[t]
\vspace*{3mm}
\centerline{%
\myfig{!}{0.7\textwidth}{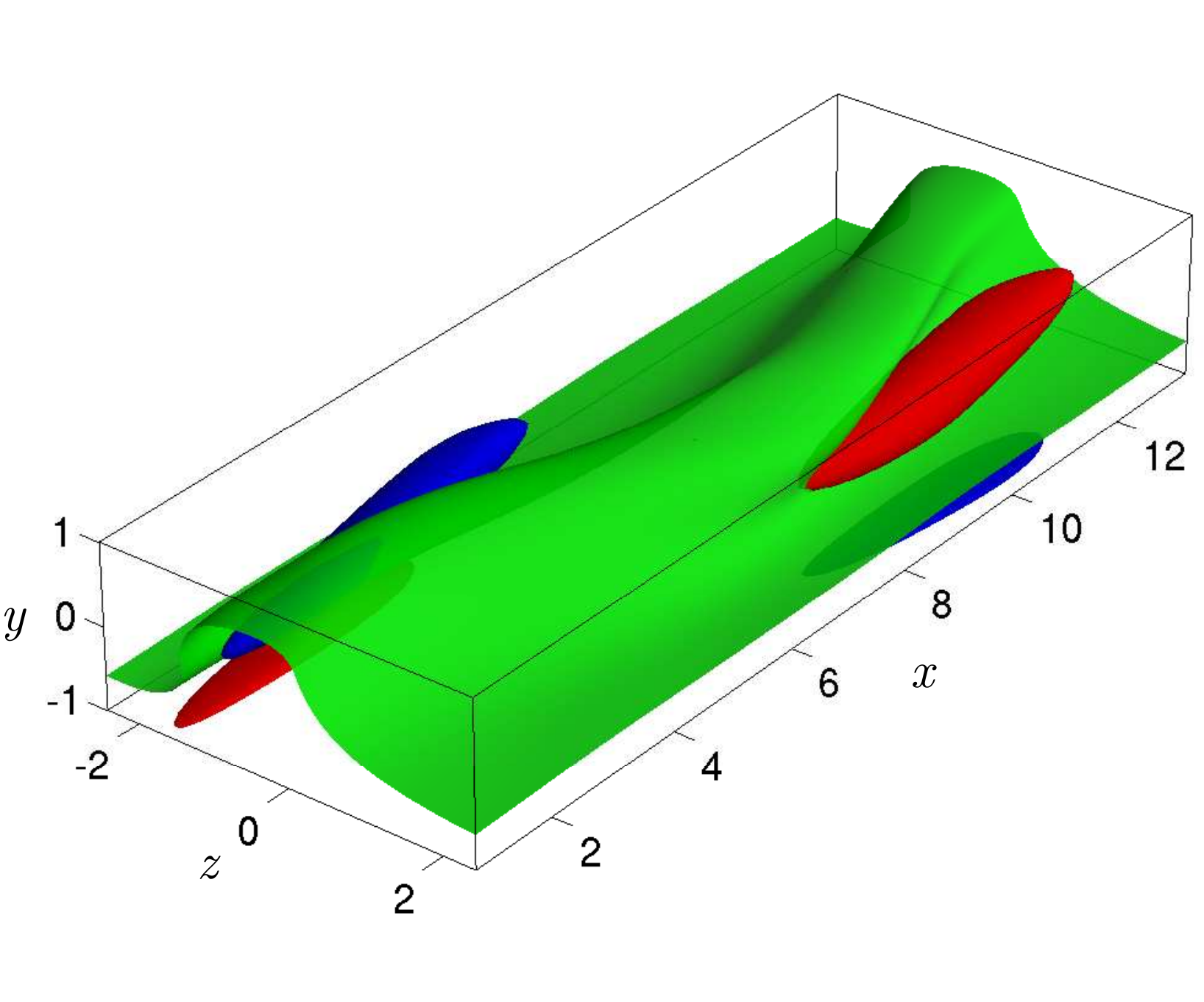}%
}
\caption[]{%
Steady travelling-wave solution
for plane Poiseuille flow, computed by Waleffe (2001, 2003).
The solution is shown at
a saddle-node
bifurcation at the lowest Reynolds number $Re_\tau\equiv u_\tau h/\nu=44$ with
the optimal streamwise and spanwise periods
$L_x^+\times L_z^+=274\times 105$.
Streamwise vortices are visualized by the red and blue isosurfaces
of the streamwise vorticity $\omega_x=\pm 0.7\max(\omega_x)$.
The corrugation of the green isosurface of the streamwise velocity
$u=0.75\max(u)$ represents a low-velocity streak.
Flow is toward positive $x$ and only half the height of the channel is shown.
}
\label{fig:waleffe}
\end{figure}

In contrast to plane Couette flow, 
laminar Poiseuille flow in a plane
channel is unstable to infinitesimal disturbances beyond a certain Reynolds
number, from where a two-dimensional equilibrium travelling-wave solution
bifurcates subcritically. A three-dimensional solution originating from
that wave was found by \cite{ehrenstein91}, while
\cite{Wally98,Wally01,Wally03} and \cite{itano01} found families of
three-dimensional steady travelling waves that are not known to be
connected to the laminar state.
In the case of
\cite{Wally98,Wally01,Wally03}, he used the homotopy approach to
construct nonlinear steady travelling waves with a reflectional symmetry
with respect to the channel central plane, which also contain streaks and
vortices, and can be continuously connected to the \cite{Nagata90}
solution for plane Couette flow \citep{Wally03}.
As shown in Figure~\ref{fig:waleffe},
the Waleffe solution for plane Poiseuille flow 
exhibits
a pair of staggered streamwise vortices and an associated wavy streak.
The solution found by \cite{itano01}
also includes streaks and vortices, but they are localized near one of the
two walls. A three-dimensional steady travelling wave with a very similar
structure was found by \cite{JimSim01} in a so-called 
`autonomous' flow
that is confined to the vicinity of one wall under the action of a damping
filter 
\citep{jpin}.

\begin{figure}[t]
\vspace*{3mm}
\centerline{%
\myfig{!}{0.4\textwidth}{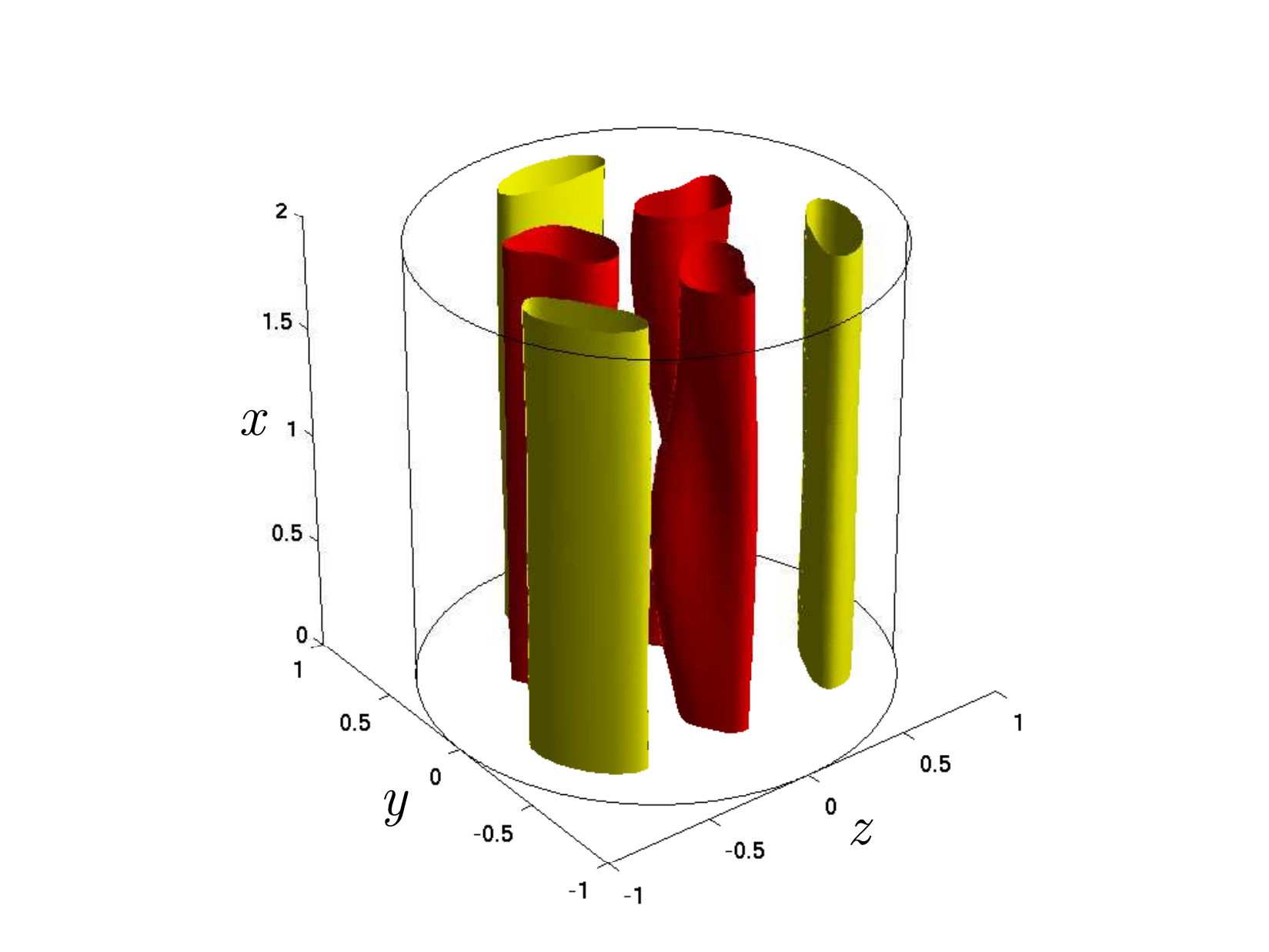}%
\mylab{-0.2\textwidth}{0.42\textwidth}{{\bf a}}
\hspace*{0.05\textwidth}
\myfig{!}{0.4\textwidth}{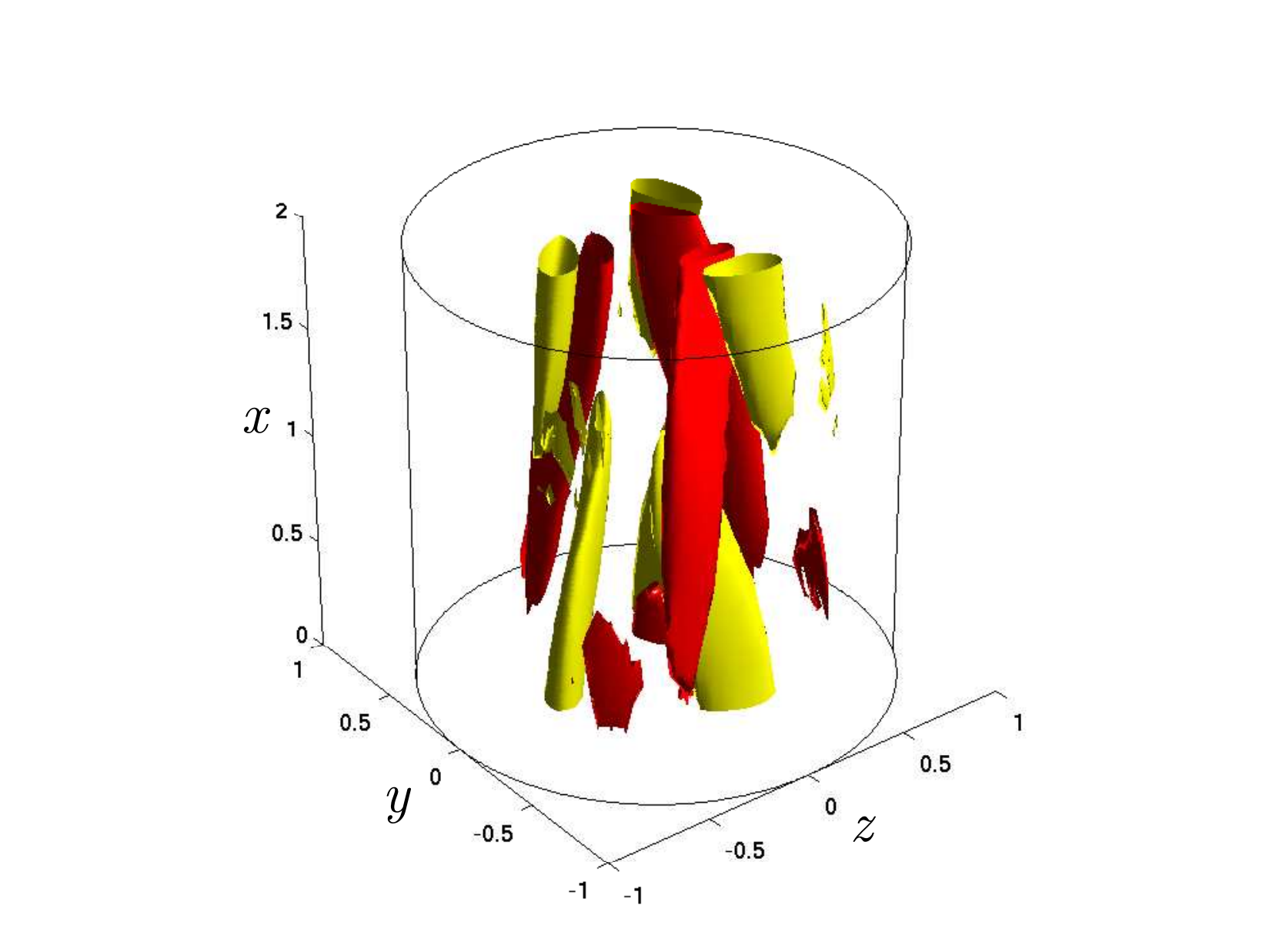}%
\mylab{-0.2\textwidth}{0.42\textwidth}{{\bf b}}
}
\caption[]{%
Perspective view of isosurfaces of
({\it a}) axial ($x$) velocity
fluctuation
$\pm 0.3U$ (yellow/red)
from laminar Hagen--Poiseuille flow and
({\it b}) axial vorticity
$\omega_x=\pm 0.6 U/D$ (yellow/red)
for
the threefold rotationally symmetric
travelling-wave solution for pipe flow
\citep{wedin04,kerswell05,kerswell07}
at Reynolds number
$Re\equiv U D/\nu=2400$,
$U$ and $D$ being
bulk mean velocity and
pipe diameter, respectively.
The axial period for the solution is
$L_x/(D/2)=2.01$.
Flow is upwards.
}
\label{fig:kerswell}
\end{figure}

The laminar Hagen--Poiseuille flow in a circular pipe is linearly
stable at all Reynolds numbers, but \cite{faisst03} and \cite{wedin04} have
recently discovered three-dimensional steady travelling-wave solutions by
using the homotopy approach proposed by \cite{Wally98,Wally03}. Both
groups obtained the same solution, which possesses discrete rotational
symmetry with respect to the pipe axis, with wavy low-velocity streaks
flanked by staggered streamwise vortices
as shown in Figure~\ref{fig:kerswell}.
The solutions originate from a saddle-node bifurcation at a finite value
of the Reynolds number.
The one with threefold rotational symmetry
appears at the lowest onset Reynolds number
\citep{wedin04},
although it was later found that travelling waves without any discrete
rotational symmetry exist at much lower Reynolds numbers
\citep{pringle07}.

As in circular pipes, the laminar flow in a square duct is linearly stable,
and no travelling-wave solutions were known until three 
such solutions
were found using the homotopy approach, introducing
either
an artificial body force or internal heating.
Two of them
found by \cite{wedin09,okino10} have streaks and streamwise vortices on only
two opposite walls, while the 
flow near the 
other two walls of the duct 
is quiescent.
The other travelling wave obtained by \cite{uhlmann10}, on the other
hand, 
features streaks and vortices in the vicinity of all four walls, 
as will be discussed in Section
\ref{second} (see Figure~\ref{fig-mu-field}).

Most of above-mentioned equilibrium solutions
for plane channels, 
circular pipes and square ducts 
look qualitatively similar
in spite of differences in their geometry and driving mechanisms
\citep{Wally98, Kawahara03}.
Their structure 
is characterized by
a wavy low-velocity streak flanked by staggered
quasi-streamwise vortices of alternating signs, resembling
the spatially-coherent objects 
found in the near-wall
region of true turbulent flows
\citep[see, e.g.][]{stretch90,jeong97}.

\subsection{Periodic Solutions}

We have just seen that three-dimensional
equilibrium solutions reproducing buffer-layer coherent structures are
available for a variety of wall-bounded flows. A smaller number of periodic
solutions have 
been reported.

In plane Couette flow, \cite{clever97} analysed the linear stability
of
the three-dimensional steady solution of \cite{Nagata90},
and they identified a Hopf
bifurcation from 
which
a three-dimensional, nonlinear
periodic solution
originates. 
They continued that solution
within its stable parameter range by forward time integration,
but found that its properties did not differ too much from those of
the steady solution due to the small amplitude of the oscillations. Using
an iterative method, 
\cite{KawKida01}
found a more unsteady three-dimensional unstable periodic solution,
which reproduces much better the full regeneration
cycle of near-wall streamwise vortices and low-velocity streaks,
as will be shown in Section~\ref{cycle}.
They also obtained a
gentler periodic solution that represents weak spanwise standing-wave
motion of the low-velocity streak.
More recently \cite{viswanath07} used
globally convergent Newton--Krylov iteration (see SIDEBAR)
to obtain five
new three-dimensional periodic solutions that demonstrate the breakup and
reorganization of near-wall coherent structures.

In plane Poiseuille flow, \cite{toh03} identified a three-dimensional
periodic-like solution that could originate from a heteroclinic connection
between two steady travelling waves that differ
from each other by a spanwise shift of half a wavelength \citep{itano01}. Their solution is
reminiscent of the heteroclinic cycle identified in a highly reduced
dynamical systems approximation to near-wall turbulence by \cite{aubry88},
who also observed a connection between two equilibria differing by a
similar spanwise shift.

More recently, \cite{duguet08} found a three-dimensional periodic
solution in Hagen--Poiseuille flow that bifurcates from the
travelling wave without any discrete rotational symmetry found by \cite{pringle07}.

Most 
unstable periodic solutions have been found for wall bounded shear
flows, but there are a few exceptions.
One is the periodic solution found
by \cite{veen06} in 
Kida--Pelz high-symmetric
flow in a triply-periodic
domain \citep{kida85}, which reproduces the \cite{kol41} universal energy
spectrum with the 
correct
energy dissipation rate
as will be shown in Section \ref{statistics},
even though its Reynolds
number is too low to identify any 
significant
inertial range. 
The inertial-range energy spectrum has not been found
in any periodic solution to the Navier--Stokes equation
yet;
however, it as well as intermittency were observed
in periodic solutions to the GOY (Gledzer--Ohkitani--Yamada) shell model
\citep{kato03}.

\subsection{Comparison of Solutions\label{comparison}}

\begin{figure}[t]
\centerline{%
%
\myfig{!}{0.5\textwidth}{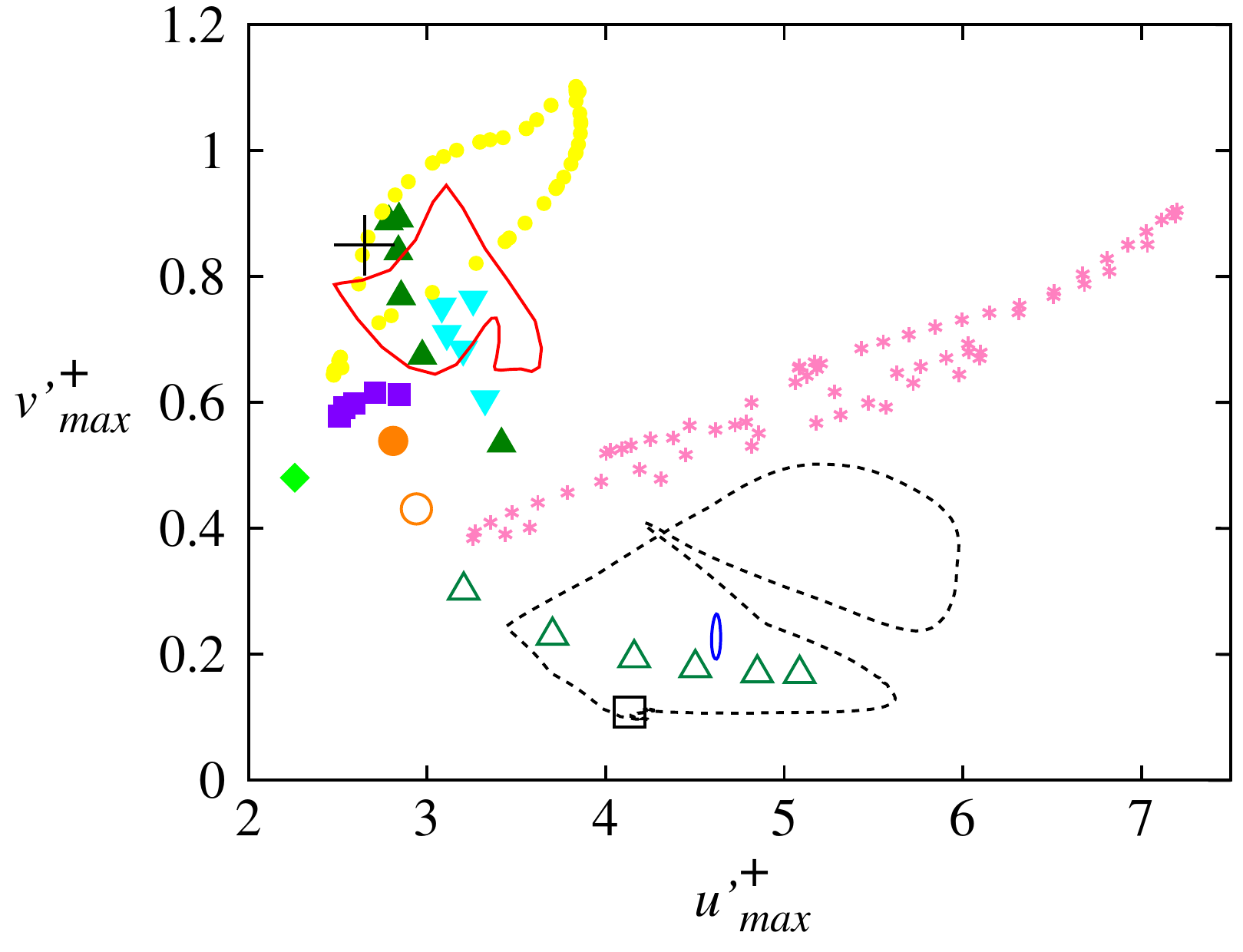}%
}
\caption[]{%
Classification into 
vortex- and streak-dominated
families of the simple invariant solutions, in
terms of their maximum streamwise and wall-normal r.m.s. velocities,
$u'_{max}$ and $v'_{max}$ \citep{jim05},
normalised with the friction velocity $u_\tau$.
Solid symbols are classified as vortex-dominated solutions,
and open ones as streak-dominated solutions.
The red and blue loops represent the dynamic ($L_x^+\times
Re_\tau\times L_z^+=190\times 34\times 130$) and the gentle ($L_x^+\times
Re_\tau\times L_z^+ =154\times 28\times 105$) periodic solutions at
$Re=400$ in \cite{KawKida01}.
{\color[named]{PineGreen}\trian}, Nagata steady solutions \citep{jim05} for several values
of the spanwise wavelength at $Re=400$ and $L_x/h=2\pi$
(upper branch, $L_z^+\times Re_\tau=76-132\times 35-35$;
lower branch, $L_z^+\times Re_\tau=53-92\times 24-25$).
{\color[named]{Purple}$\blacksquare$}, autonomous solutions in \cite{JimSim01}
($L_x^+\times L_z^+\times \delta^+=151-189\times 180\times 38-46$,
where $\delta$ is the filter height).
{\color[named]{YellowOrange}{\Large \circle}}, \cite{Wally03} travelling waves
(upper branch,
$Q/\nu=1303$, where $Q$ is the volume flux per unit span,
$L_x^+\times Re_\tau\times L_z^+= 387\times 123\times 149$;
lower branch,
$Q/\nu=1390$, $L_x^+\times Re_\tau\times L_z^+ = 379\times 121\times 146$).
$\square$, \cite{itano01} asymmetric wave for $Q/\nu=4000$,
$L_x/h\times L_z/h=\pi\times 0.4\pi$.
The dotted loop is the periodic-like solution of \cite{toh03}.
{\color[named]{SkyBlue}$\blacktriangledown$}, temporal averages of \cite{viswanath07} periodic solutions
(cases $P_2-P_6$ in his table~1) for $Re=400$ and $L_x/h\times
L_z/h=1.75\pi\times 1.2\pi$.
{\color{green}$\blacklozenge$}, upper branch of \cite{uhlmann10} eight-vortex
travelling wave ($Re=1371$, $L_x/h=2\pi$).
{\color[named]{Yellow}\solidcircle}, \cite{wedin04,kerswell05,kerswell07}
fourfold rotationally symmetric travelling waves for several values
of the streamwise period at $Re=2000$
(${\cal R}_4$ travelling waves in figure 8 of \cite{kerswell05}).
{\color[named]{Lavender}$*$}, 
twofold rotationally symmetric travelling waves
\citep{wedin04,kerswell05,kerswell07}
for several values
of the streamwise period at $Re=2000$
(${\cal R}_2$ travelling waves in figure 8 of \cite{kerswell05}).
%
We have also shown the turbulent state computed by \cite{kmm}
($L_x^+\times Re_\tau\times L_z^+=2300\times 180\times 1150$)
by the symbol {\Large $+$} for comparison.
}
\label{compare}
\end{figure}

As reviewed above,
equilibrium and periodic solutions reproduce buffer-layer coherent structures
in near-wall turbulence at least qualitatively.
The characteristics of those solutions are
now quantitatively
compared in Figure~\ref{compare},
which has been adapted from \cite{jim05}  by adding some of the solutions that
have been found since then. Each solution is represented by a single point
whose coordinates are the maximum values of the streamwise and wall-normal
velocity fluctuation, $u'_{max}$ and $v'_{max}$, of its
intensity profiles.
The streamwise and wall-normal velocity fluctuations have
been taken as
representing the intensities of the streamwise velocity streaks and of the
quasi-streamwise vortices, respectively.
Despite differences in their boundary conditions
most solutions can be seen to fall into one of two classes: a
``vortex-dominated''
family, characterized by smaller $u'_{max}$ and
larger $v'_{max}$, and a 
``streak-dominated''
one, characterized by
larger $u'_{max}$ and smaller $v'_{max}$.

In the upper-left corner of Figure \ref{compare} we find the vortex-dominated
solutions, and in the lower-right corner the streak-dominated ones. The
former are represented
by \cite{Nagata90} upper-branch steady solutions and
the \cite{KawKida01} dynamic periodic solution for Couette flow,
by \cite{JimSim01} autonomous travelling waves and
the \cite{Wally03} upper-branch travelling wave for Poiseuille flow,
and by \cite{wedin04,kerswell05,kerswell07}
fourfold rotationally symmetric travelling waves
in pipe flow.
\cite{viswanath07} periodic solutions in plane Couette flow are also part
of this family.
The vortex-dominated invariant solutions 
are able to 
reproduce the velocity
fluctuations comparable with near-wall turbulence.

The streak-dominated family includes
\cite{Nagata90} lower-branch steady solutions and
the \cite{KawKida01} gentle periodic solution for Couette flow, and
the \cite{Wally03} lower-branch travelling wave
for Poiseuille flow,
which, 
however, 
is too close to the saddle-node bifurcation to differ 
significantly 
from
the corresponding upper branch.
The steady travelling wave obtained by \cite{itano01}
and the heteroclinic connection identified by \cite{toh03}
in Poiseuille flow are also in the streak-dominated family.
The streak-dominated solution is probably not directly related to turbulence,
and it is now believed to play an important role in the subcritical
transition to turbulence \citep[see][]{kerswell05, eckhardt07, eckhardt08}.
At subcritical Reynolds numbers
some of the streak-dominated solutions for plane Poiseuille and Couette
flow have only one unstable direction in phase space, so that those
invariant sets and their stable manifolds form the laminar-turbulent
separatrix \citep{itano01,Wally03,kawahara05,wang07,schneider08}.

Although closer to the streak-dominated family, 
the twofold rotationally symmetric travelling waves in pipe flow
found by \cite{wedin04,kerswell05,kerswell07} 
seem to be in between the two families.

The upper-branch of 
the eight-vortex travelling wave in a
square duct 
found by \cite{uhlmann10}
has weak velocity fluctuations, but
is relatively close to the vortex-dominated solutions.
The corresponding lower-branch solution
(not shown),
on the other hand,
has a weaker
$u'_{max}$ than the upper-branch one, and is thus quite different from the
streak-dominated solutions.
However,
such weak fluctuations in a square duct
are believed to be a consequence of averaging only in the
streamwise direction.

\section{SIGNIFICANCE OF INVARIANT SOLUTIONS\label{significance}}

In this section
we shall discuss the possible relevance of the simple invariant solutions
to turbulent flows.
All the solutions described above are unstable
at the Reynolds numbers at which turbulence is observed, but the dimensions
of their unstable manifolds in phase space are typically low
\citep[see][]{Wally01,kawa3,kawahara05,wang07,veen06,kerswell07,viswanath09}. Therefore,
although we should not expect to observe them as such in real
turbulence, a generic turbulent solution could approach them and spend a
substantial fraction of its lifetime in their neighbourhood. 
Each of the invariant solutions has stable and unstable manifolds,
and their intersections contain connecting orbits for the state of the
system to follow as it visits these building blocks, associated with 
coherent structures. Some of these connecting orbits are associated
with chaotic dynamics described mathematically by Smale's horseshoe
map \citep{wigg}. Such structures are called homoclinic and heteroclinic
tangles. Hence, stable and unstable manifolds of the
above simple invariant solutions could represent turbulence dynamics, while
the simple solutions themselves would represent coherent structures
embedded in a turbulent state.

\subsection{Near-Wall Regeneration Cycle\label{cycle}}

In the buffer region of wall turbulence
it is known that
there exist self-sustaining coherent structures,
namely quasi-streamwise vortices and low-velocity streaks,
which control the near-wall flow dynamics.
Low-velocity streaks have been recognized to be
generated by cross-streamwise
advection of vorticity lines of
mean shear flow by streamwise vortices
\citep{kline67},
and it is believed that the streamwise vortices arise from
the instability of the streak and the subsequent
nonlinear vortex stretching in the streamwise direction
\citep{Hamilton95,Schoppa02,Kawahara03}.
These mutual generation processes are thought to form a closed nonlinear
regeneration cycle of the streaks and the vortices
\citep{Hamilton95,Waleffe97}.
Such a nonlinear cycle (in other words, self-sustaining process)
of buffer-layer coherent structures
has been described theoretically
in terms of equilibrium or temporally periodic
solutions to the incompressible Navier--Stokes
equation for plane channel flows.

In order to relate a nonlinear equilibrium state
with the self-sustaining process of
buffer-layer coherent structures
\cite{Wally01, Wally03} introduced a spanwise periodic
array of streamwise-independent rolls
as external forcing to the Navier--Stokes equation.
In this forced system there is a steady state that is
represented by the low- and high-velocity streaks
formed under the action of the imposed
streamwise vortices.
This basic steady state, composed of streaks,
can lose its stability to a sinuous disturbance
beyond some critical value of the roll strength, 
i.e. the amplitude of the external forcing.
The nonlinear three-dimensional steady (travelling-wave)
solutions, i.e.\ Nagata's solution in
plane Couette flow and Waleffe's
solution
in plane Poiseuille flow \citep{Nagata90,Wally01,Wally03},
bifurcate from the basic state subcritically,
yielding the zero-forcing
state which
is a solution to the Navier--Stokes equation.
Such nonlinear solutions can be considered to
represent the self-sustaining process
in the sense that
they appear from the formation of the streaks by
the vortices and the generation of the vortices
through the streak instability.

\begin{figure}[t]
%
\centerline{
\myfig{!}{0.25\textwidth}{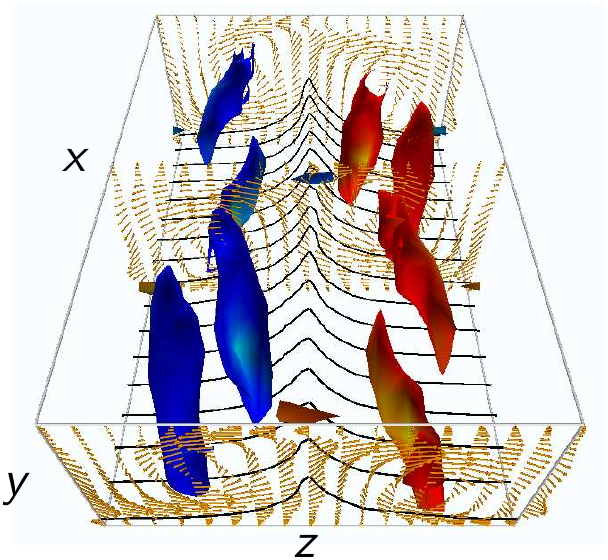}
\myfig{!}{0.25\textwidth}{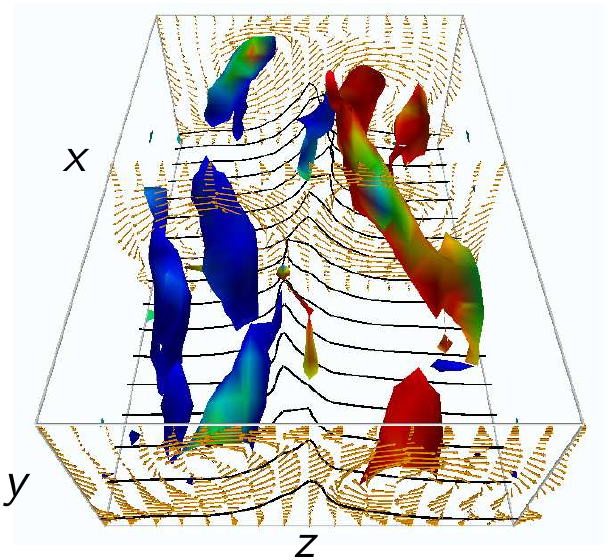}
\mylab{-0.29\textwidth}{0.22\textwidth}{{\bf a}}
}
%
\centerline{
\myfig{!}{0.25\textwidth}{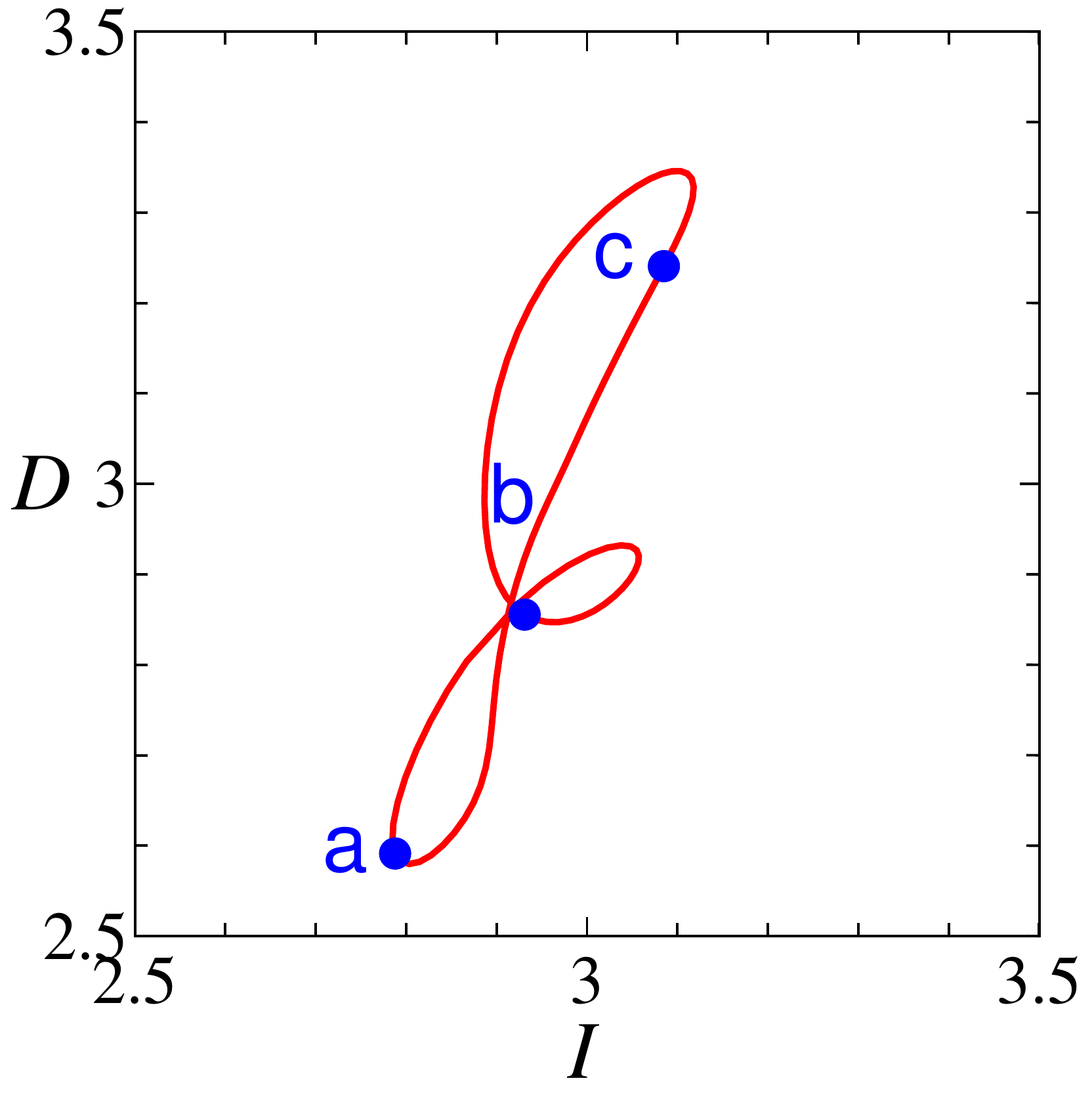}
\myfig{!}{0.25\textwidth}{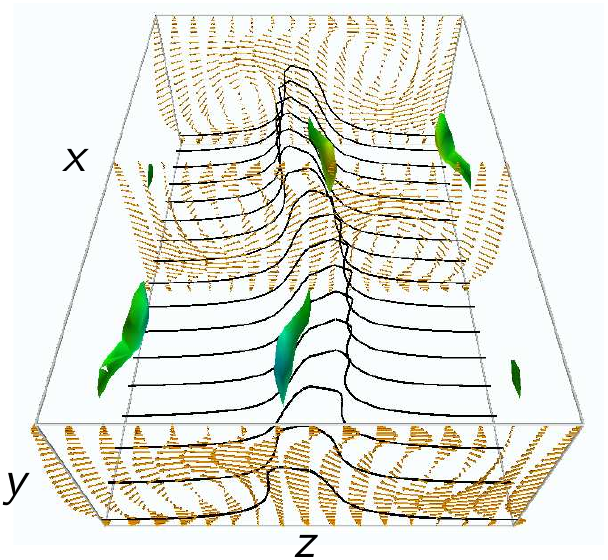}
\myfig{!}{0.25\textwidth}{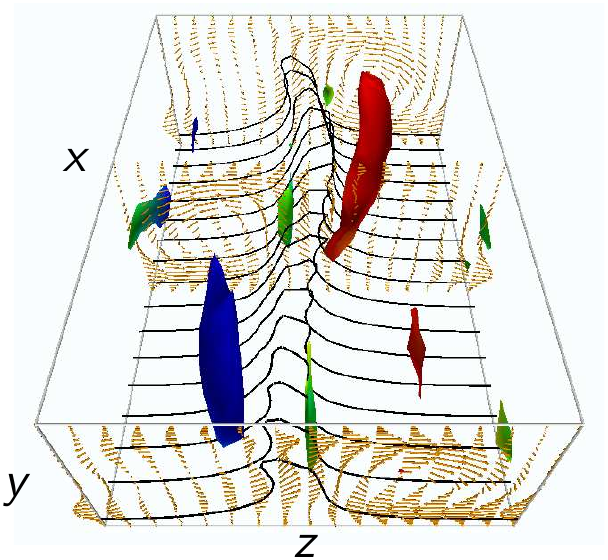}
\mylab{-0.29\textwidth}{0.22\textwidth}{{\bf b}}
\myfig{!}{0.25\textwidth}{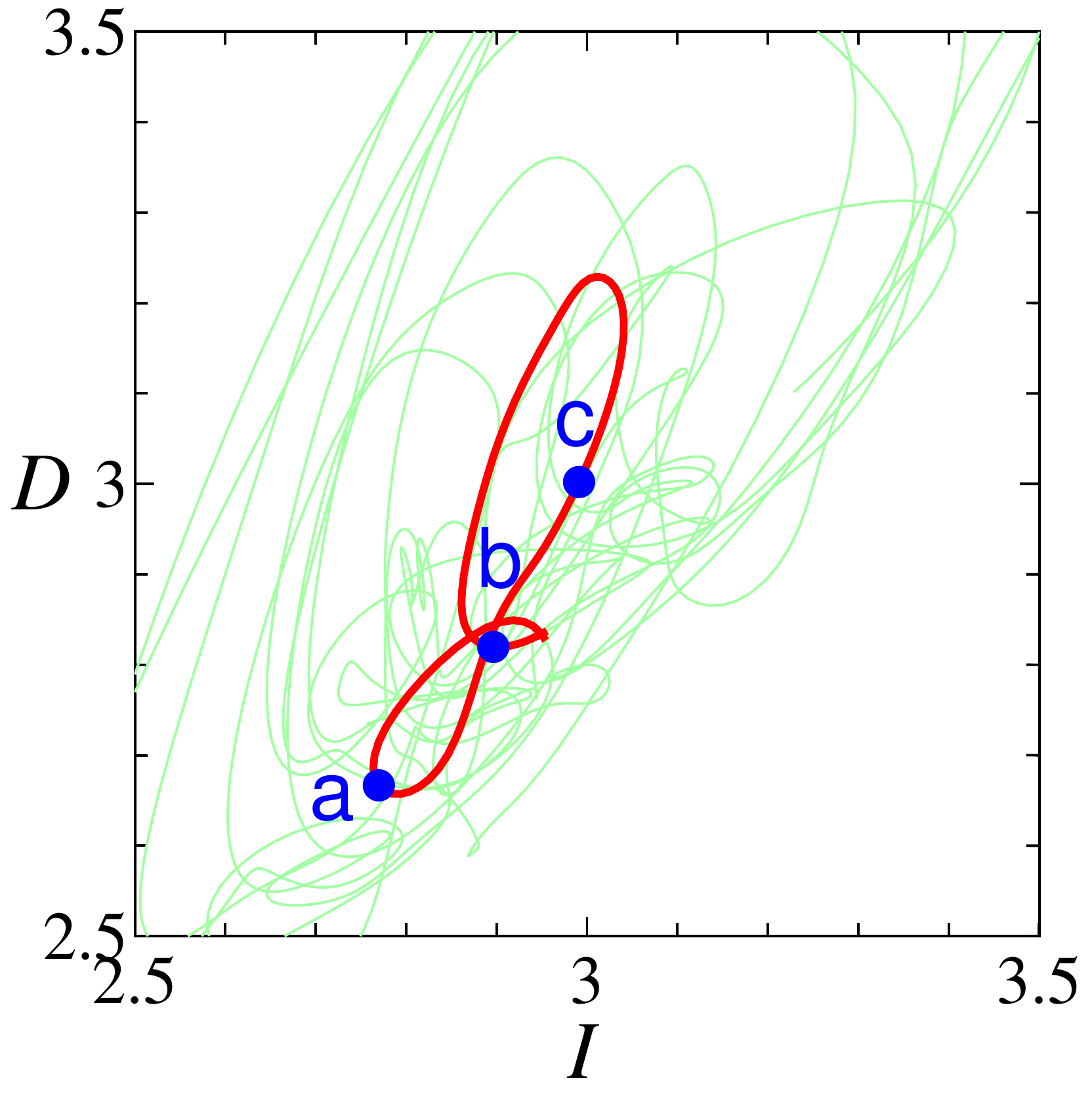}
}
%
\centerline{
\myfig{!}{0.25\textwidth}{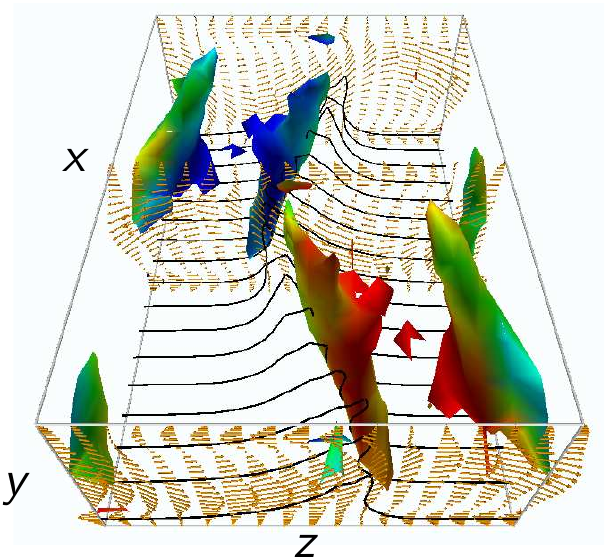}
\myfig{!}{0.25\textwidth}{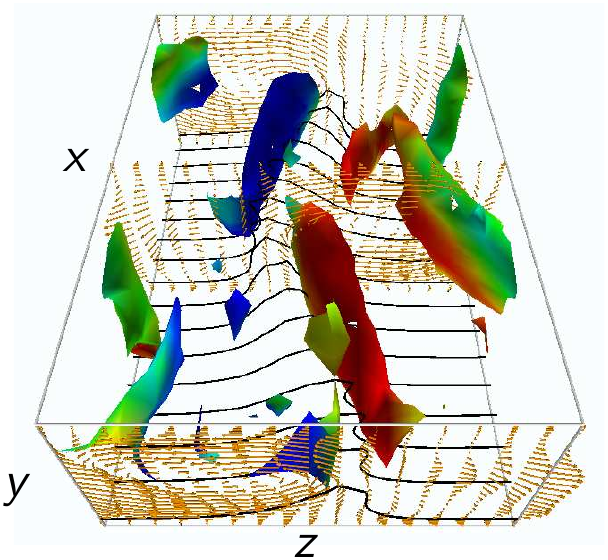}
\mylab{-0.29\textwidth}{0.22\textwidth}{{\bf c}}
}
\caption[]
{
A cycle of evolution of spatial structures of the \cite{KawKida01}
vortex-dominated time-periodic solution (left)
and those of the turbulent solution (right) at corresponding phases
for plane Couette flow at $Re=400$.
Flow structures are visualized in the whole spatially periodic box 
($L_x\times 2h\times L_z$) over one full cycle at three times 
with a constant interval $21.6 h/U$.
The upper (or lower) wall moves into (or out of) the page
at velocity $U$ (or $-U$).
Vortex structures are represented by
isosurfaces of the Laplacian of pressure, 
$\nabla^2 p=0.15\rho (U/h)^2$,
where $\rho$ is the mass density of the fluid.
Color of the isosurfaces of $\nabla^2 p$ indicates the 
sign of the
streamwise ($x$) vorticity: red is positive (clockwise),
blue is negative
(counter-clockwise).
Cross-flow velocity vectors and
contours of the streamwise velocity 
at $u=-0.3 U$ are also shown on cross-flow planes $x=$ const.
The streamwise and spanwise period for the solution are
$L_x/h\times L_z/h =1.755\pi\times 1.2\pi$.
The phases for the visualization are identified by
the symbols {\bf a}, {\bf b}, {\bf c} in
the phase plane $(I, D)$ respectively for
the periodic (left) and the turbulent (right) solution.
$I$ and $D$ are the total energy input and dissipation
normalized with those of a laminar state.
The segment of the turbulent (green) orbit
exhibiting an approach to the periodic solution
is colored red.
}
\label{figperistruc}
\end{figure}

The nonlinear equilibrium solutions are considered to be the
simplest description of the self-sustaining process
of near-wall coherent structures.
In order to describe the dynamics of their regeneration cycle,
however, we should obtain more complex solutions.
One candidate of such solutions is the vortex-dominated periodic solution
found by \cite{KawKida01} for plane Couette flow.
A full cycle of the temporal evolution of spatial structures of
this periodic solution
is depicted in Figure~\ref{figperistruc} ({\it a}--{\it c}). 
Spatial structures of a turbulent solution
are also shown at corresponding phases.
The friction Reynolds number of the periodic solution
is $Re_\tau = 34$, and the time period is $65h/U$
($=188\nu/u_\tau^2$).
The streamwise and spanwise computational periods are
$L_x^+\times L_z^+ = 188\times 128$.
The dynamics of the periodic solution
is described by a cyclic sequence of events which consists of
\begin{enumerate}
\item[(i)] Figure~\ref{figperistruc} ({\it a}): 
the formation of low-velocity streaks
through the cross-streamwise advection of streamwise velocity
induced by decaying streamwise vortices;
\item[(ii)] Figure~\ref{figperistruc} ({\it b}): 
the development of the streaks, and
their subsequent sinuous bending along the streamwise direction as well as tilting in 
the spanwise ($z$) direction; and
\item[(iii)] Figure~\ref{figperistruc} ({\it c}): 
the regeneration of streamwise vortices as the consequence
of the above streak instability in
Figure~\ref{figperistruc} ({\it b}).
\end{enumerate}
This cyclic sequence
is completely consistent with a previously reported
regeneration cycle
\citep{Hamilton95, Waleffe97}, and
the wavy low-velocity streak and staggered counter-rotating
streamwise vortices at phase ({\it b})
are closely similar to the coherent structures
commonly observed in the buffer layer
turbulence
\citep{stretch90,jeong97}.

\subsection{Turbulence Statistics for the Viscous Range\label{statistics}}

There are two types of canonical turbulent flows,
one of which is wall turbulence and the
other of which is isotropic turbulence.
In the near-wall region
of any turbulent flow,
which is referred to as the inner layer,
the mean velocity is observed to scale with
the kinematic viscosity
$\nu$ and the wall friction
velocity $u_\tau$
\citep{prandtl}.
This is known as the Prandtl wall law, and
one of the most typical universal statistical laws
of fluid turbulence.
It is also expected that statistical properties
of small-scale turbulence in
any flow away from a wall
are nearly isotropic as a consequence of
a nonlinear energy cascade process.
The energy spectrum of small-scale motion
in such flows is known to scale with $\nu$
and the mean energy dissipation rate per unit mass $\epsilon$
\citep{kol41}.
This is what we call the Kolmogorov similarity law, 
another universal law of turbulence.
Here we discuss 
the relevance of the invariant solutions
to turbulence statistics.

\begin{figure}[t]
\centerline{
\myfig{!}{0.45\textwidth}{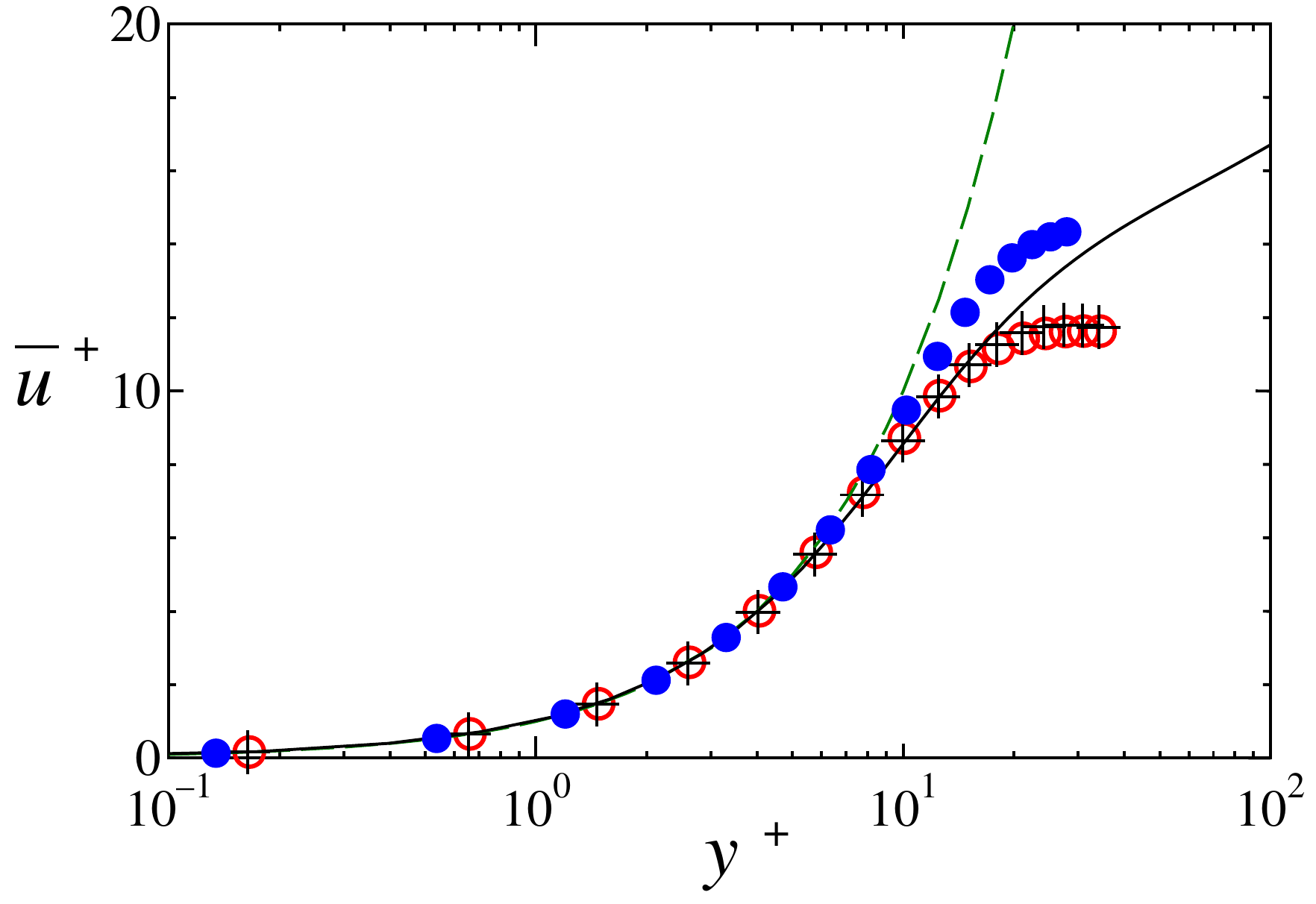}
}
\caption[]
{
Comparison of the normalised mean velocity profiles
between the \cite{KawKida01} time-periodic and turbulent flows.
The mean streamwise velocity $\overline{u}^+=\overline{u}/u_\tau$ is shown
as a function of the distance from the wall $y^+=yu_\tau/\nu$.
{\color{red}{\Large \circle}},
the vortex-dominated periodic solution at $Re_\tau=34$ representing regeneration cycle;
{\color{blue}{\Large \solidcircle}},
the streak-dominated gentle periodic solution at $Re_\tau=28$.
+, the turbulent solution for the same condition;
\solid, the turbulent solution
for plane Poiseuille flow at $Re_\tau=590$
\citep{moser99}.
{\color[named]{PineGreen}\dashed}, the laminar solution $\overline{u}^+=y^+$.
}
\label{figmean}
\end{figure}

Figures~\ref{figmean}
compares the normalised mean velocity
profile for the \cite{KawKida01}
time-periodic solutions in plane Couette flow
with those for turbulent flow.
Open red symbols denote the vortex-dominated
periodic solution ($Re_\tau=34$; see the large red loop in Figure~\ref{compare}) representing
the regeneration cycle, while closed blue symbols indicate
the streak-dominated
periodic solution with a gentle variation
($Re_\tau=28$; see the small blue loop in Figure~\ref{compare}).
The mean velocity for the vortex-dominated periodic solution is 
in excellent agreement with that for the turbulent solutions
not only at $Re_\tau=34$ but also at $Re_\tau=590$
\citep{moser99} within the buffer layer,
say at $y^+ \lesssim 20$, while
that for the streak-dominated solution is closer to
the laminar velocity profile.
Although the logarithmic velocity profile in the overlap region cannot be
seen in the vortex-dominated solution at the low Reynolds number
$Re_\tau=34$, this solution reproduces
Prandtl's scaling law with $\nu$ and $u_\tau$
in the viscous wall layer.

\begin{figure}[t]
\centerline
{
\myfig{!}{0.7\textwidth}{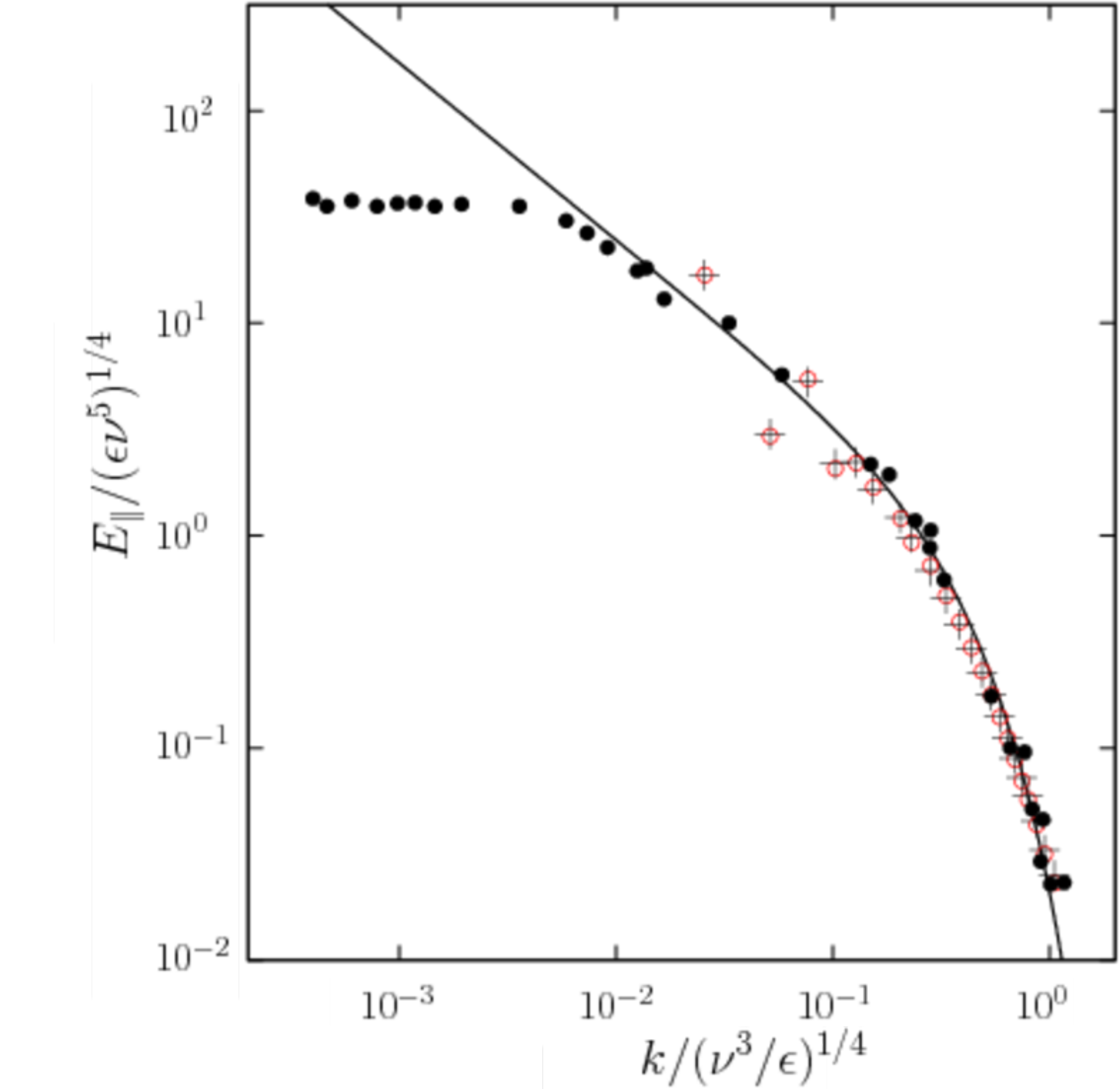}
}
\caption[]
{
One-dimensional longitudinal energy spectrum $E_\parallel$
for the \cite{veen06} periodic solution in Kida--Pelz high-symmetric flow. 
The lateral and longitudinal
axes are normalised
by $(\nu^3/\epsilon)^{1/4}$ and $(\epsilon\nu^5)^{1/4}$,
respectively.
{\color{red}{\Large \circle}}, the periodic state of high-symmetric flow at $R_{\lambda}=67$; 
+, the turbulent state of high-symmetric flow;
\solidcircle, experimental data of homogeneous shear turbulence
at $R_{\lambda}=130$ 
taken from \cite{Champagne70}.
\solid, the high-$R_{\lambda}$ asymptotic form derived theoretically by the sparse
direct-interaction approximation \citep{Kida97}.
}
\label{spectrum}
\end{figure}

In Figure~\ref{spectrum}, 
the normalised energy spectrum
for a periodic orbit obtained by \cite{veen06} in Kida--Pelz
flow \citep{kida85}
at the Taylor-microscale Reynolds number
$Re_\lambda=67$ is shown, along with the spectrum
for the turbulent solution.
The time period of the solution is
$58 (\nu/\epsilon)^{1/2}$.
The spectra for the periodic and turbulent solutions are
in excellent agreement, and they are consistent
with the turbulence spectra obtained by
experiments
\citep{Champagne70} and by 
closure theory
\citep{Kida97}
in the high-wavenumber range,
implying that the periodic solution
represents the Kolmogorov dissipation-range energy
spectrum.
Actually,
it has been confirmed that around $Re_\lambda=67$
the energy dissipation rate
$\epsilon$ for the periodic solution
is nearly independent of $\nu$ and
takes almost the same value as that of 
turbulence.
This result also implies that the energy spectrum
for the periodic
solution scales with $\nu$ and $\epsilon$ as
for the turbulent state.
As shown in Figure~\ref{spectrum}, however,
the Reynolds number, at which the periodic solution
has been obtained, is not high enough to clearly
show the $-5/3$ power spectrum for the inertial range
observed both in the experiments and the theory.

%
%

It is curious that averaging along a single periodic solution of
a relatively short period
for plane Couette flow \citep{KawKida01}
and Kida--Pelz flow \citep{veen06}, reproduces so
well typical turbulence statistics for the viscous range, such as the
mean velocity profile in the buffer layer and
the dissipation-range energy spectrum,
which are long-time averages along true turbulent
orbits.

One interesting idea was proposed by \cite{kawasaki}. They conjectured that 
relatively short periodic solutions give good statistics for quantities such as the
energy dissipation rate because these are averages over the spatial domain.
In a relatively large spatial domain, many quasi-independent processes
of 
flow dynamics
will occur simultaneously and thus averaging over space is
not unlike averaging over time.

\citet{saiki09} studied periodic solutions
to the Kuramoto--Sivashinsky equation. They compared time-mean values along the periodic
solutions to those along chaotic segments of equal length, and found that the former
are more narrowly distributed around the same average. This observation suggests that
a good representation of flow statistics might be a generic property of relatively short
periodic solutions embedded in spatio-temporal chaos.

\subsection{Mean Secondary Flow \label{second}}

Turbulent flow through straight pipes with non-circular cross-section
exhibits a mean secondary flow in the cross-sectional plane, whereas
laminar flow does not. A satisfactory description of the mechanism
leading to the formation of such secondary flow of Prandtl's second
kind has been an outstanding issue for many decades, despite
continuous efforts from a number of researchers.  
In the following we will focus our attention on one geometry,
namely the flow through a square duct.  
Recent vortex eduction studies based on DNS data have shown that the 
mean secondary flow can be linked to a non-homogeneous probability
distribution of coherent flow structures along the duct perimeter
\citep{uhlmann07,pinelli10}.  
In a statistical sense, mean secondary flow vorticity can be
understood as a footprint of the preferential spatial occurrence of
streamwise vortices caused by the cross-sectional geometry. 
These DNS studies have also revealed a succession of the most probable
instantaneous streak-vortex patterns when increasing the Reynolds
number from a value just above the limit for sustained turbulence
($Re\equiv Uh/\nu=1100$, $U$ being the average mean
velocity and $h$ the duct half-width) to moderate values ($Re=3500$). 
At the lowest Reynolds numbers, the near-wall regeneration cycle is
often  simultaneously active at only two opposite walls (both
featuring a single low-velocity streak and corresponding streamwise
vortices), while the flow near the other pair of walls is quiescent
(regime I).  
Above $Re\approx1400$, the most probable state is characterized by
simultaneous turbulence activity
(a single streak associated with vortices)
on all
four walls (regime II). 
When further increasing the Reynolds number above $Re\approx2000$, the 
most probable number of low-velocity streaks per wall increases to two
(regime III).  
Further transitions have not been described in detail, due to the
limited range of Reynolds numbers accessible to DNS. 
The statistical mean flow, on the other hand, always features the same
number of two secondary flow vortices per wall (i.e.\ a total of eight
per cross-section), with a given sense of rotation, albeit with
changing shape as a function of the Reynolds number.

Among the presently known travelling waves, the
solutions of \cite{okino10} and \cite{uhlmann10} 
have the potential to represent coherent structures in turbulent duct
flow in different respective Reynolds number ranges. 
The former solution exhibits coherent structures which appear to be
consistent with observations in turbulent flow pertaining to regime I,
while the latter solution matches some of the characteristics
related with coherent structures
found in regime II
as well as the mean secondary flow in both regimes II and III,
as will be discussed next.

\begin{figure}[t]
  \begin{center}
    \begin{minipage}{.75\linewidth}
      \includegraphics[width=\linewidth]
      {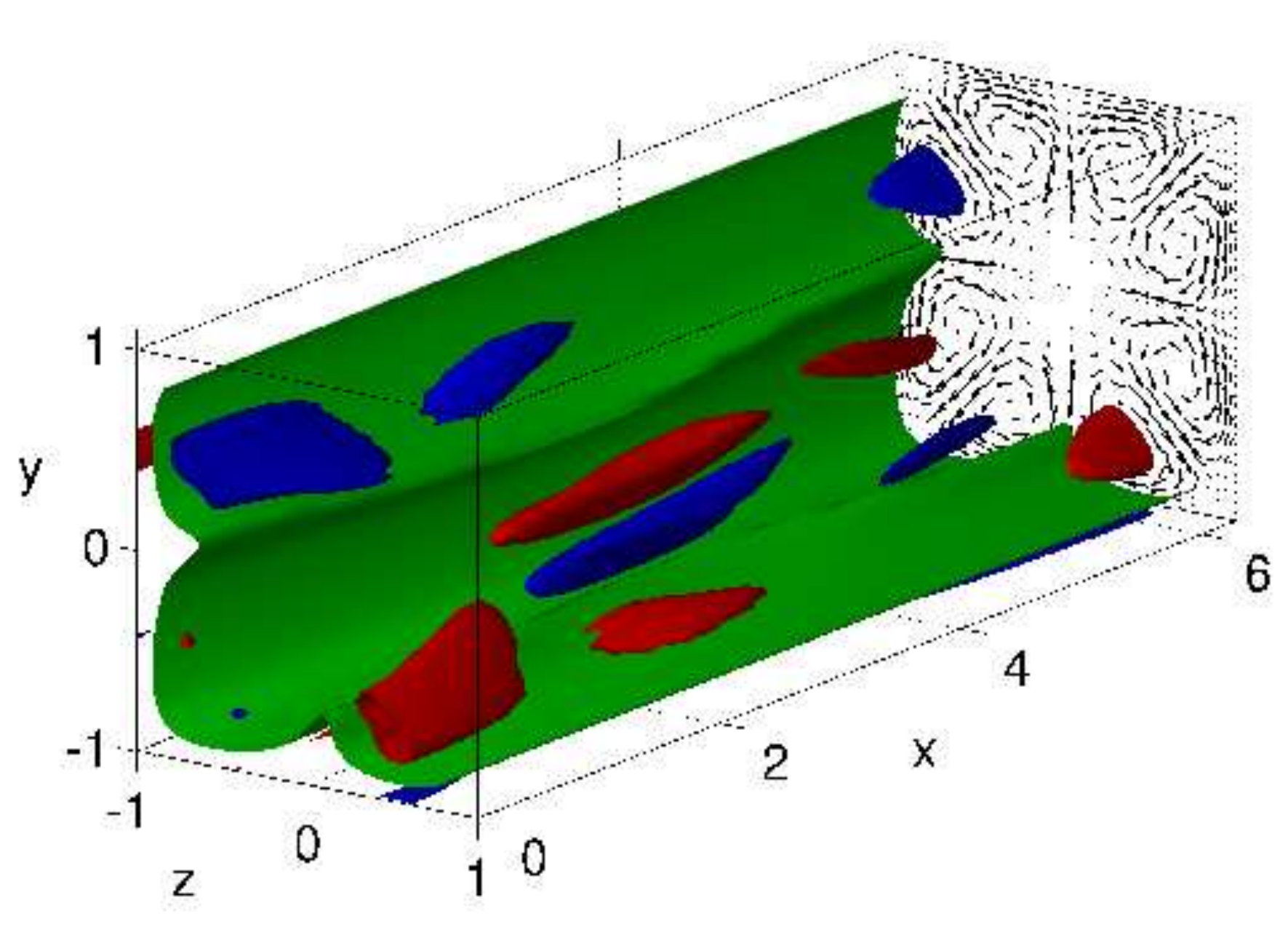}
    \end{minipage}
  \end{center}
  \caption{ 
    Travelling wave in pressure-driven
    square-duct flow \citep{uhlmann10}.
    The upper-branch solution at Reynolds number $Re\equiv Uh/\nu=1404$
    ($U$ and $h$ being the bulk mean velocity and the duct half-width, respectively)
    and streamwise period $L_x/h=2\pi$
    is shown by means of surfaces of constant
    values of the total streamwise velocity $u=0.55\max(u)$ (sheet-like
    structure near the walls, green color) 
    and of the streamwise vorticity at $\pm0.65\max(\omega_x)$
    (tubular structures, blue/red color). 
    The surfaces are cut away on one side of a
    diagonal of the duct cross-section for improved clarity. 
    Mean secondary flow vectors are projected upon the 
    back plane of the graph. 
    Flow is toward positive $x$.
  }
  \label{fig-mu-field}
\end{figure}

\begin{figure}[t]
  \begin{center}
    \begin{minipage}{5ex}
      $\displaystyle
      \frac{E_{vw}^{1/2}}{U}$
    \end{minipage}
    \begin{minipage}{.6\linewidth}
      \includegraphics[width=\linewidth]
      {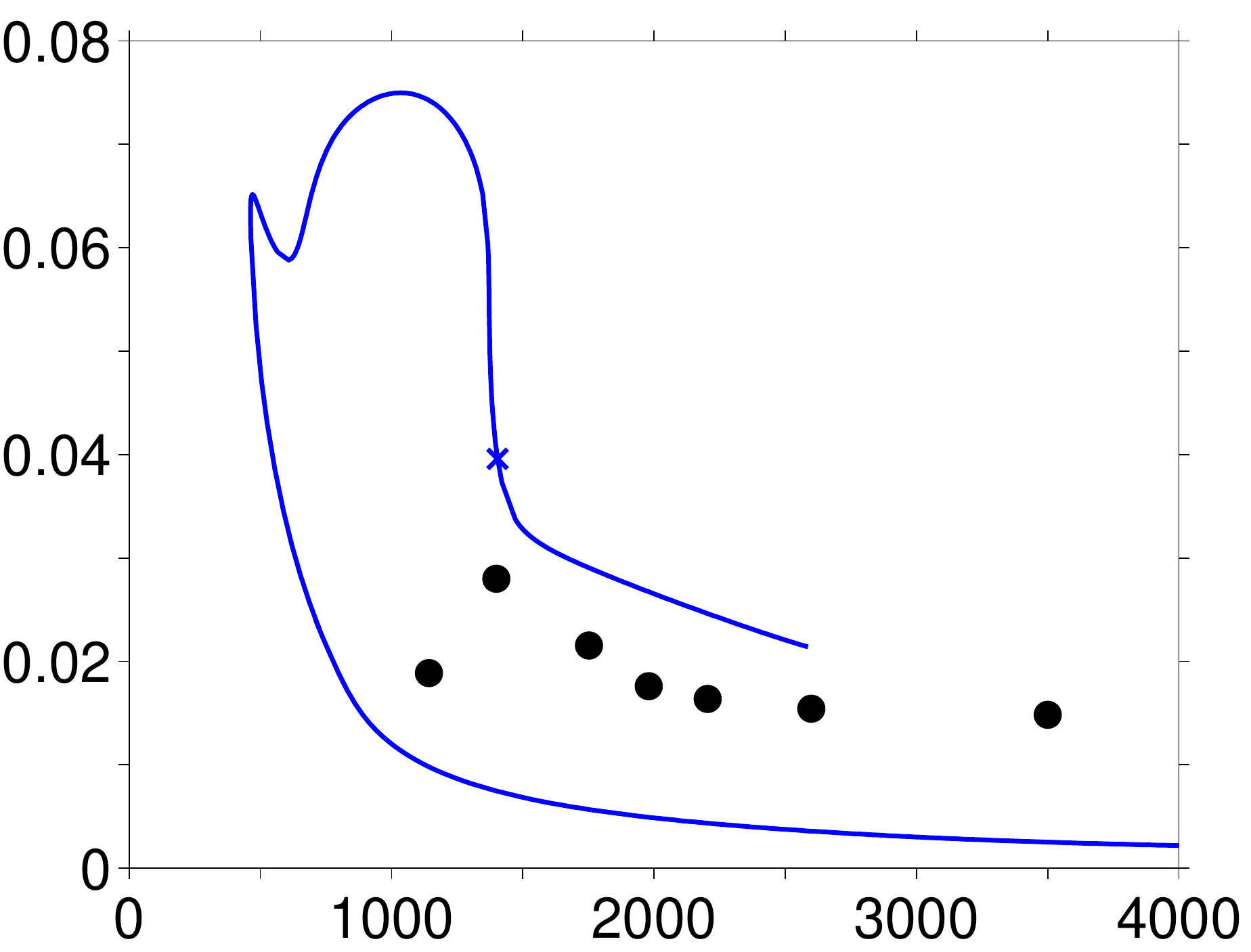}
      \centerline{$Re$}
    \end{minipage}
  \end{center}
  \caption{ 
    The intensity of the mean secondary flow $E_{vw}^{1/2}$ 
    (where $E_{vw}$ is the energy of the streamwise-averaged flow in
    the cross-section, defined in an integral sense) 
    for the duct flow travelling waves of \cite{uhlmann10}, 
    normalized by the bulk velocity $U$ and shown as a function of
    the Reynolds number. 
    The blue solid line ({\color{blue}\solid}) connects solutions
    at constant streamwise 
    period $L_x/h=2\pi$;
    the cross
    ({\color{blue}$\boldsymbol{\times}$}) 
    corresponds to 
    the flow field visualized in figure~\ref{fig-mu-field}.
    The closed symbols ($\bullet$) are for long-time averaged turbulent
    flow in domains with length $L_x/h\approx4\pi$
    \citep{uhlmann07}. 
  }
  \label{fig-mu-enervw}
\end{figure}

Focusing on the solution by \cite{uhlmann10},
Figure~\ref{fig-mu-field} shows the 
flow field on the upper branch at $Re=1404$.
It can be seen that a wavy low-speed streak as well as staggered
streamwise vortices are found in the vicinity of all four wall
planes, as 
observed
in turbulent flow in regime II. 
Furthermore, the streamwise average of this solution exhibits a 
secondary flow pattern with eight vortices, strikingly resembling the 
statistically averaged secondary flow observed in turbulence at
the same Reynolds number \citep{uhlmann10}. 
The comparison of this travelling wave solution with turbulence data is
further elaborated in Figure~\ref{fig-mu-enervw}, where the secondary
flow intensity 
is shown as a function of the bulk Reynolds number. 
At low Reynolds numbers (pertaining to regime I) the upper branch
solution of \cite{uhlmann10} by far exceeds intensity levels found in
turbulent flow;  
beyond $Re\approx1400$ (i.e.\ in regimes II and III), however, both quantities
are comparable,
implying that the upper branch could represent mean secondary flow at any value of
higher Reynolds number.
In contrast with the secondary flow the
perturbation energy as well as the friction factor (not shown) 
are 
found to match statistical data from turbulence in a narrow
range of Reynolds numbers corresponding to regime II. 

In conclusion, we state that it is significant for simple invariant solutions to
represent not only spatio-temporal coherence in turbulent flows but also
typical statistical properties of turbulence.


\section{FUTURE DIRECTIONS\label{future}}

Finally, we briefly discuss possible future areas of
research in this nascent field of the relation 
of
the
coherent structures and turbulence statistics 
to
dynamical systems. Although it has
turned out that the structures, the dynamics, and the statistics of
buffer-layer coherent structures are tantalizingly close to those of some
nonlinear equilibria or periodic solutions, the dynamical interpretation of the
large-amplitude intermittent events observed in the near-wall region is
still unclear. The invariant solutions computed so far are not very useful
for explaining intermittent behaviour
in turbulence
because they lack localization.
They have been computed in small, periodic, domains
and persist for all time.

In order to describe the intermittency
we will need to compute solutions which display localization in the spatial
domain, the temporal domain or both. Good candidates for localized solutions
are homoclinic and heteroclinic orbits. In the temporal domain, i.e. in the
sense of 
Section \ref{connect}, such solutions correspond to large
excursions in phase space while transiting between simple invariant solutions.
In the spatial domain, such solutions correspond to different phases of the
flow coexisting in physical space, glued together by sharp interfaces.
Computation of orbits connecting invariant sets in time is hard because
they need to be approximated over long time intervals. The sensitive dependence
on initial conditions makes this computation potentially unstable, as
any kind of numerical noise is amplified during time-stepping.

\begin{figure}[t]
\centerline{
\myfig{!}{0.7\textwidth}{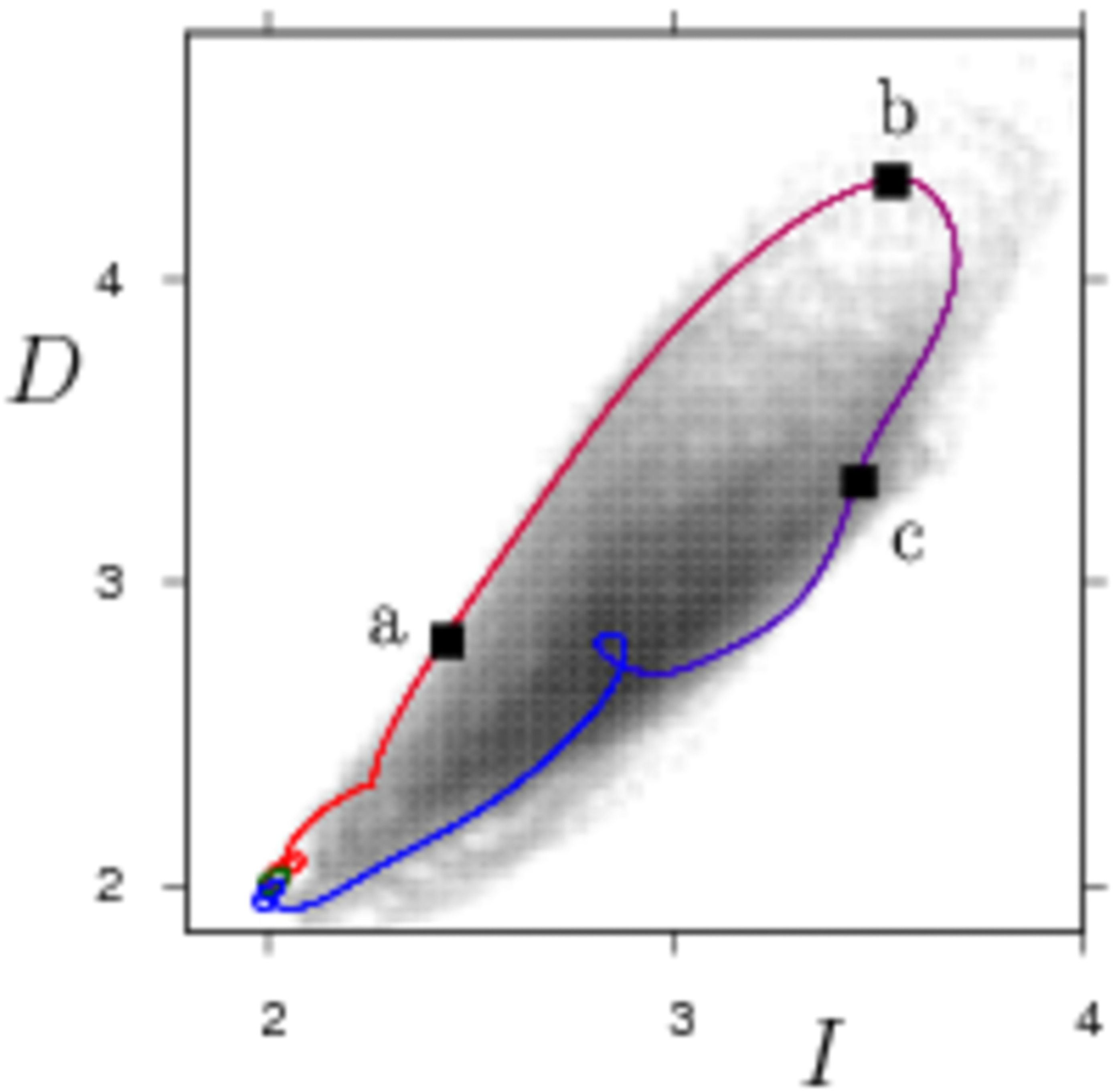}
}
\caption[]
{
Projection of the orbit homoclinic to the gentle periodic solution (shown in green)
in plane Couette flow at $Re=400$
onto the energy input $I$ and the dissipation rate $D$,
normalised by their value in laminar flow \citep{veen11}.
The piece of orbit leaving the gentle periodic solution is shown in red and the one approaching it in
blue.
In the background, the PDF (Probability Density Function) of transient turbulence is shown in gray scale.
The labels {\bf a}--{\bf c} correspond to the snap shots in Figure \ref{fighomostruc}.
}
\label{fighomo}
\end{figure}

\begin{figure}[t]
\centerline{
\myfig{!}{0.3\textwidth}{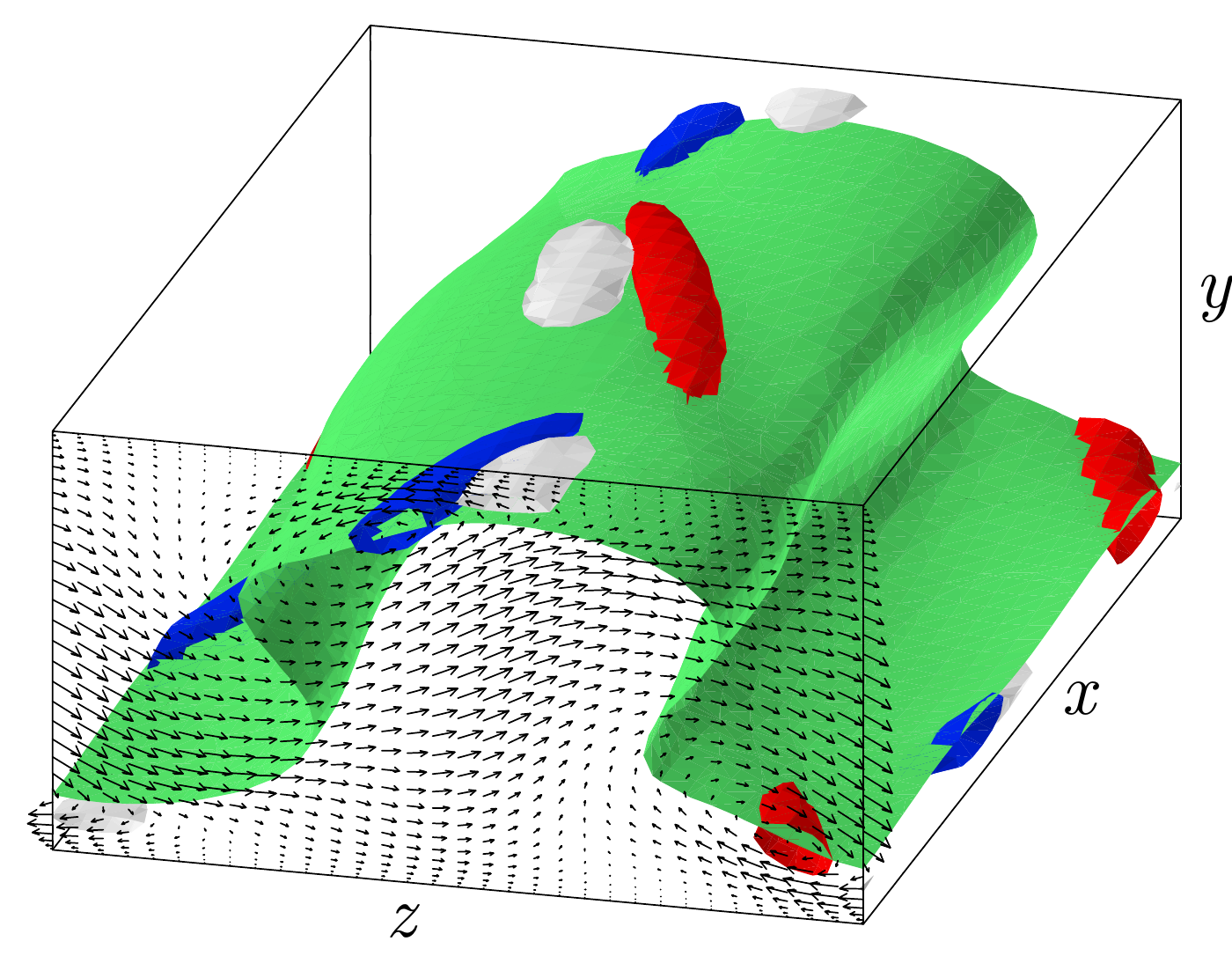}
\myfig{!}{0.3\textwidth}{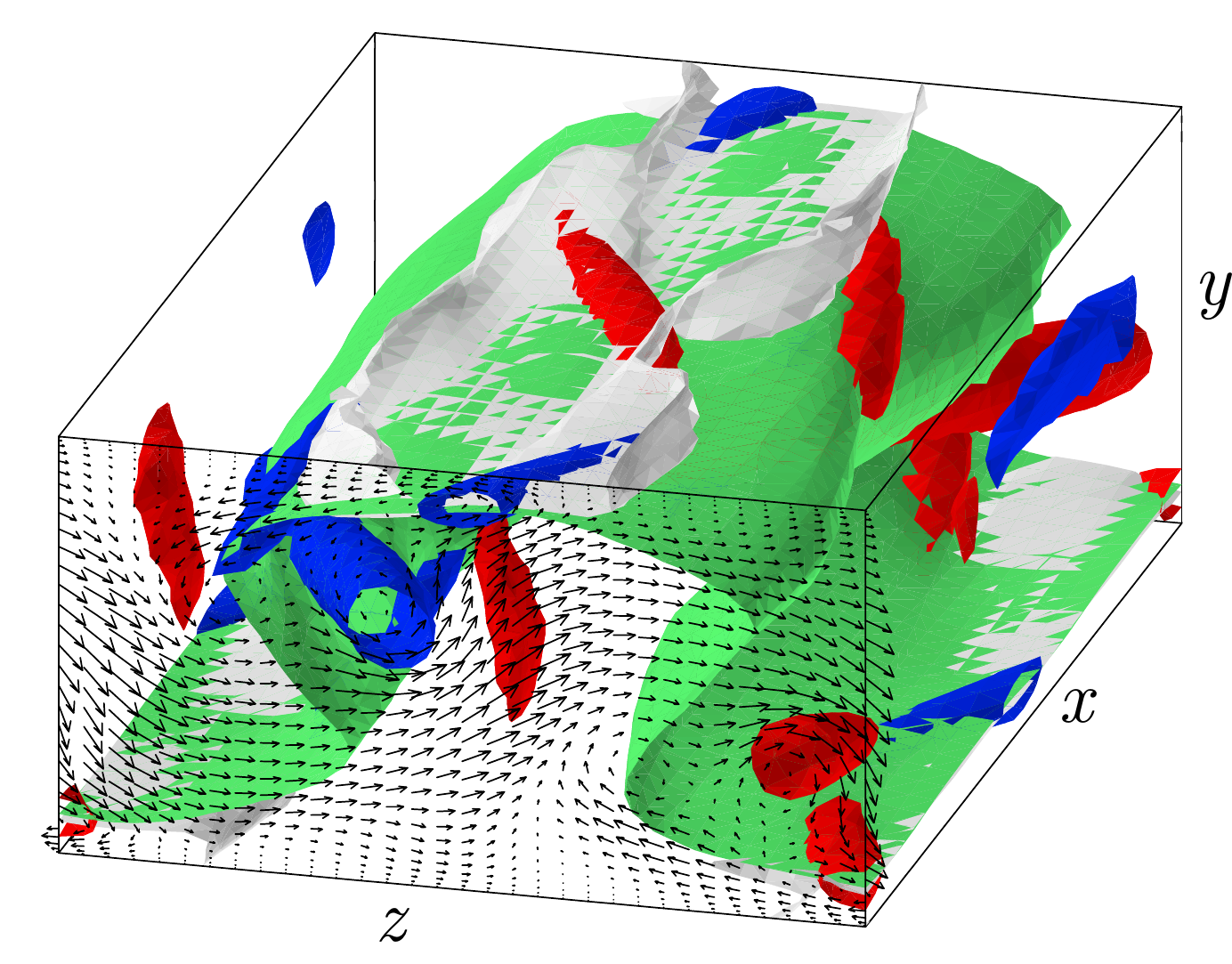}
\myfig{!}{0.3\textwidth}{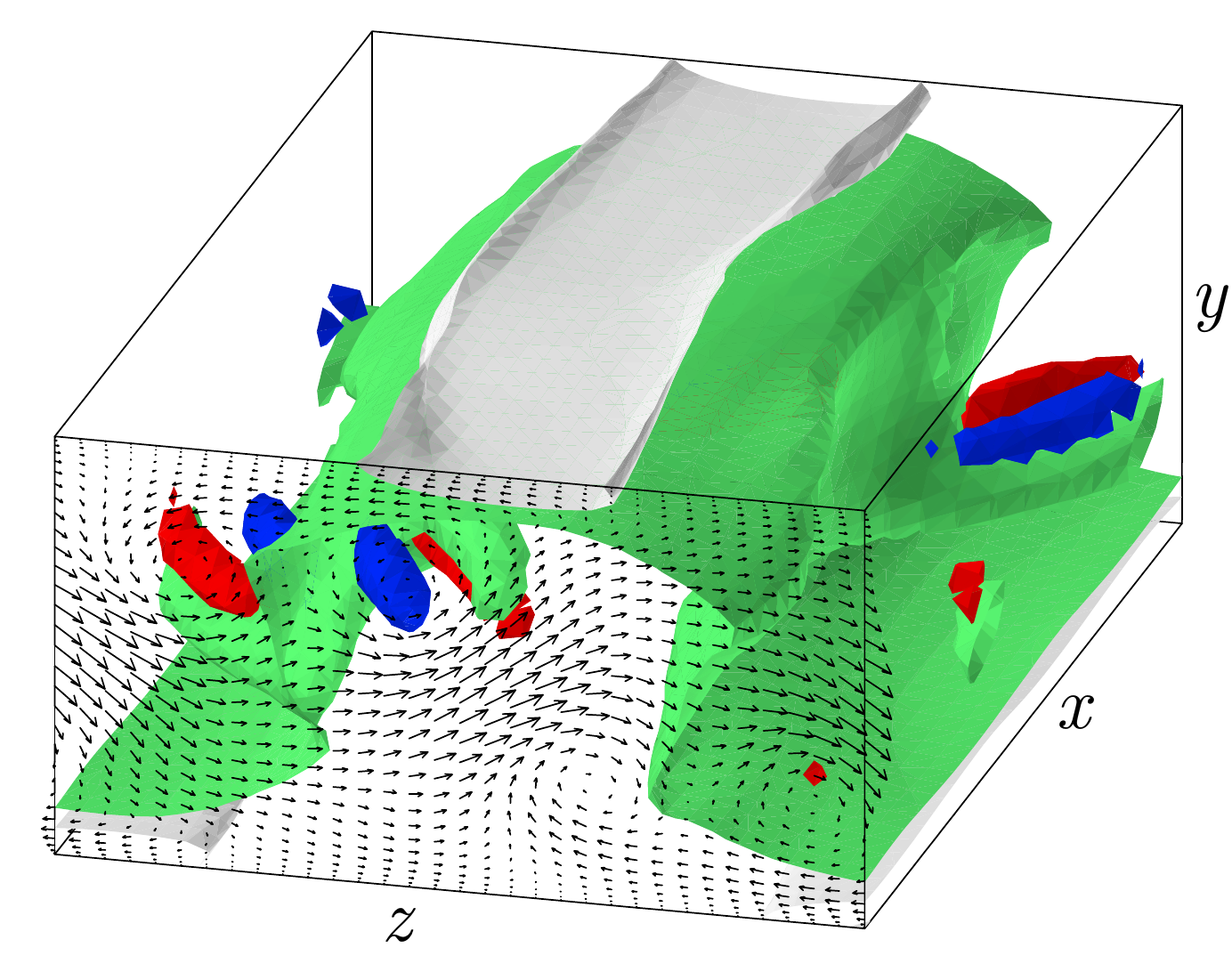}
\mylab{-0.98\textwidth}{0.32\textwidth}{{\bf a}}
\mylab{-0.58\textwidth}{0.32\textwidth}{{\bf b}}
\mylab{-0.2\textwidth}{0.32\textwidth}{{\bf c}}
}
\caption[]
{
Visualizations of flow structures in one periodic box $L_x\times 2h\times L_z$
at three phases on the orbit homoclinic to the gentle periodic solution
in plane Couette flow at $Re=400$ \citep{veen11},
labeled as in Figure~\ref{fighomo}.
The green, corrugated isosurfaces of the null streamwise ($x$) velocity
represent a streamwise streak.
Red and blue objects are isosurfaces of $0.4\rho(U/h)^2$
for the Laplacian of the pressure,
and denote the vortex tubes of the positive and negative streamwise-vorticity component.
Gray isosurfaces show the local energy dissipation at $20$ times the value of
laminar dissipation.
Cross-stream ($y,z$) velocity is also shown in the plane $x=0$.
The flow symmetry suggests that cross-stream velocity in the plane
$x/h=L_x/2$ is given by the reflection of that in the plane $x=0$ in the spanwise ($z$)
direction.
}
\label{fighomostruc}
\end{figure}

An attempt to explain bursting, i.e. temporal intermittency, through connecting
orbits was made
recently
by \citet{veen11}. They studied the
gentle periodic solution
identified by \citet{KawKida01}. This periodic solution has a single unstable multiplier
and its stable manifold separates the phase space into two parts, one containing the
domain of attraction of the laminar flow \citep{kawahara05}. A piece of the two-dimensional
unstable manifold can
be computed by Newton--Krylov continuation of orbit segments. A boundary
value problem is set up which specifies that the initial point of the segment
must lie in the linear approximation of the manifold, whereas the final point satisfies
some scalar condition, e.g. the energy dissipation rate has a fixed value.
This boundary value problem is 
under-determined 
by a single degree of freedom and
thus we can compute a family of orbit segments by arclength continuation \citep{veen10}.
Since a connecting orbit separates the manifold into two components, the
consecutive orbits segments can converge to a connecting orbit during the
continuation.
In Figure \ref{fighomo} an orbit homoclinic
to the gentle periodic solution is shown, which makes a large excursion during which the 
energy dissipation rate reaches more than twice the mean value in periodic
motion and four times the value at the laminar equilibrium. 
In the dissipative phase
of this bursting cycle, as shown in Figure \ref{fighomostruc},
at the valley and the crest of the streaks
small quasi-streamwise vortices are formed and move rapidly in the
spanwise direction, and the larger part of the energy dissipation takes
place there.
Similar dissipative flow structures are also observed during bursting in
a turbulent state, and they
are very different from those for
the vortex-dominated periodic solution,
in other words the regeneration cycle, in Figure \ref{figperistruc}.

For now only a few spatially localized invariant solutions are available.
Such solutions were first computed in plane Couette flow by \cite{cherhabili97}
and revisited by \cite{ehrenstein08}. They found three-dimensional equilibrium
states which are localized in the streamwise direction.
\cite{schneider10} studied the same flow at lower Reynolds numbers, but focused
on the dependence of certain equilibria and traveling waves
on the spanwise extent of the computational domain. They found that for 
domain sizes substantially larger than the minimal flow unit, some of these
equilibria become localized in the spanwise direction. Towards the edges of the spawise domain, 
the flow is laminar but in the centre there are a number of streamwise vortices.
The creation of vortex pairs takes place at saddle-node bifurcations in an
intricate diagram called a homoclinic snake.
Further search for localized invariant solutions should lead to
understanding of
typical spatially-intermittent turbulence, such as
turbulent spots \citep{dauchot95}, oblique stripes \citep{prigent02} and puffs
\citep{wygnanski73}.

Another important problem to be tackled is
the 
relevance of simple invariant solutions 
to 
high-Reynolds-number
turbulent flows.
As reviewed in this paper,
essential dynamics of buffer-layer structures has
been described in terms of invariant solutions.
They can also reproduce
the
mean velocity profile in the near-wall layer of turbulent flow and
the energy spectrum in the dissipation range of isotropic turbulence.
However, we have no simple invariant solutions which
reproduce the universal statistical laws
of turbulence, i.e.\ the logarithmic velocity profile or
the $-5/3$-power energy spectrum in high-Reynolds-number turbulence.
If such solutions exist, there would be an interesting
question as to
whether or not they could
represent larger-scale structures and their dynamics in the logarithmic layer
\citep[see][]{jim11} and in the inertial range.
\cite{flores10}
have recently found quasi-periodic bursts in the logarithmic region of fully
turbulent channel flow, strongly reminiscent of the regeneration cycle in the buffer layer,
suggesting that the large-scale dynamics in the logarithmic layer might be related to invariant
solutions in larger periodic domains at high Reynolds number.
This problem is quite challenging to a dynamical systems approach
to turbulent flows as well as to
large-scale scientific computations.


\section*{SUMMARY POINTS}

\begin{enumerate}
\item
An overview of a dynamical systems approach to turbulence has been presented, based on recently found simple invariant
solutions and connecting orbits in Navier-Stokes flow.
\item
In wall-bounded shear flows, these solutions come in two families: vortex-dominated and
streak-dominated. The former family reproduces
buffer-layer coherent structures and velocity fluctuations comparable to near-wall turbulence,
while the latter appears to regulate transition to turbulence in the presence of a stable laminar flow.
\item
Vortex-dominated periodic solutions in plane Couette flow have been shown to reproduce
the regeneration cycle of the coherent structures and Prandtl's wall law
in the viscous wall layer,
while periodic solutions 
in Kida--Pelz flow
have been found with a Kolmogorov spectrum in the
dissipation range. 
\item
In the square duct, travelling-wave solutions have been found
which
lead to consistent patterns and intensities as compared to
turbulence-driven mean secondary flow.
\item
In plane Couette flow,
equilibria and travelling waves have been computed which exhibit localization either in the spanwise
or the streamwise direction. Also, a connecting orbit has been computed which shows 
turbulent bursting
localized in time. Such solutions reproduce the intermittent nature of transitional flows.
\end{enumerate}

\section*{FUTURE ISSUES}

\begin{enumerate}
\item
Can we
obtain spatially-localized invariant solutions to reproduce
spatial intermittency of turbulence, such as
turbulent spots, puffs and stripes observed in large domains?
\item
Can we 
find
periodic solutions in sufficiently large domains and at sufficiently large Reynolds number
to reproduce the logarithmic
law for wall bounded flows and Kolmogorov's intertial-range scaling for isotropic turbulence?
\item
Can we compute a sufficient number of invariant solutions for a given flow to
be fully decomposed into building blocks, i.e.\ simple invariant solutions
(coherent structures) and their connecting orbits (turbulence dynamics)?
\end{enumerate}

\section*{ACKNOWLEDGMENTS}
The authors cordially thank Prof. F. Waleffe, Prof. R. Kerswell and Prof. D. Viswanath
for their original figures and data for
Figures \ref{fig:waleffe}, \ref{fig:kerswell} and \ref{compare}.
Several of the results presented in Sections \ref{review} and \ref{significance}
are outcomes from the authors' collaboration with
Prof. S. Kida, Prof. J. Jim\'enez, Prof. M. Nagata and Dr. A. Pinelli.
The authors greatly appreciate their discussions in the course of the collaboration.
G.K. is supported by a Grant-in-Aid
for Scientific Research (B) from the Japanese Society for the Promotion of Science.
L.v.V. is supported by NSERC Grant nr. 355849-2008.

\bibliography{kawahara,veen}
\end{document}